\documentclass{iopjournal}
\usepackage{amsmath}

\begin{document}

\articletype{Paper}

\title{A Particle-in-Cell Simulation Framework for Thomson Scattering Analysis in Inertial Confinement Fusion}

\author{Zi'ang Zhu$^1$\orcid{0009-0001-2362-4698}, Yifan Liu$^1$\orcid{0009-0002-9014-1298}, Jun Li$^{1,2,*}$\orcid{0000-0001-9247-0760}, Han Wen$^3$\orcid{0000-0001-8567-380X}, Shihui Cao$^4$\orcid{0000-0003-1066-3708}, Yin Shi$^1$\orcid{0000-0001-9902-873X}, Qing Jia$^1$\orcid{0000-0001-5051-9925}, Chaoxin Chen$^4$\orcid{0000-0003-1228-2327}, Yaoyuan Liu$^4$\orcid{0000-0002-8846-1527}, Hang Zhao$^4$\orcid{0000-0002-2356-367X}, Tao Gong$^4$\orcid{0000-0001-6885-6962}, Zhichao Li$^{4,*}$\orcid{0000-0002-8379-6889}, Dong Yang$^4$\orcid{0000-0003-4245-7204} and Jian Zheng$^{1,2}$\orcid{0000-0001-8290-4772}}

\affil{$^1$Department of Plasma Physics and Fusion Engineering and CAS Key Laboratory of Frontier Physics in Controlled Nuclear Fusion, University of Science and Technology of China, Hefei, Anhui 230026, China.}

\affil{$^2$Collaborative Innovation Center of IFSA (CICIFSA), Shanghai Jiao Tong University, Shanghai 200240, China.}

\affil{$^3$School of Physics and Electronics, Hunan University, Hunan, Changsha 410082, China.}

\affil{$^4$National Key Laboratory of Plasma Physics, Laser Fusion Research Center, China Academy of Engineering Physics, Sichuan, Mianyang 621900, China.}

\affil{$^*$Author to whom any correspondence should be addressed.}

\email{junlisu@ustc.edu.cn and lizhi@mail.ustc.edu.cn}

\keywords{particle-in-cell simulation, inertial fusion energy}

\begin{abstract}
In inertial confinement fusion (ICF), Thomson scattering (TS) is a widely used diagnostic technique for probing plasma conditions. We present a first-principles numerical approach to obtaining scattered light signals of ion acoustic features with high resolution in angle and frequency space using particle-in-cell simulations under typical ICF conditions. Our method demonstrates good agreement with existing theories for thermal collective TS. In the super-thermal collective regime, the results align with theory when the driven plasma modes are well-matched in wave vectors to the probe and collecting beams. Moreover, we also find that TS signals can remain significant even under imperfect wave-vector matching—a result that contradicts the conventional expectation that the TS spectrum strictly follows the plasma density spectrum. We attribute this discrepancy to a beating wave mechanism arising from the interaction between the probe beam and driven plasma density modulations. Our work thus provides a practical framework for interpreting TS signals from driven ion modes, a common yet complex feature in ICF plasmas.
\end{abstract}

\section{INTRODUCTION}
In the pursuit of ignition and high-gain inertial confinement fusion (ICF), accurate diagnosis of high-energy-density (HED) plasmas generated by the interaction of intense laser beams with spherical targets or hohlraums is essential.  
Within these HED plasmas, a variety of physical processes can arise that complicate diagnostics, including the Langdon effect \cite{langdon1980,turnbull2020,turnbull2024}, electron thermal transport \cite{luciani1983,henchen2019}, and laser–plasma instabilities (LPIs)\cite{michel2009,kirkwood2013,craxton2015,li2023,wang2023,hao2023,liu2024,lian2025}. 
These processes commonly occur in nonequilibrium or collisional HED plasmas and significantly influence the efficiency of laser energy coupling.

Thomson scattering (TS), the elastic scattering of low-energy photons by electrons, has become a well-established diagnostic technique for probing HED plasma conditions, driven waves and distribution functions in ICF.
Collective Thomson scattering (CTS) arises when the scattered waves from individual electrons interfere coherently, allowing access to information about the collective behavior of the plasma.\cite{sheffield2011}

The power of the scattered light of CTS, denoted as $P_s(\vec{R},\omega_s)$, reflects characteristics of electron density perturbations. The scattered power in the unit frequency $d\omega_s$ and unit solid angle $d\Omega$ is given by\cite{sheffield2011}:
\begin{equation}
P_s(\vec{R},\omega_s)d\Omega d\omega_s=\frac{P_ir_0^2}{A2\pi}d\Omega d\omega_s\left(1+\frac{2\omega}{\omega_i}\right)|\hat{s}\times(\hat{s}\times\hat{E}_{io})|^2NS(\vec{k},\omega),
\end{equation}
where \( P_i \) denotes the power of the incident (probe) beam, \( r_0 \) is the classical electron radius, and \( A \) is the cross-sectional area of the scattering volume. The unit vector \(\hat{s}\) denotes the direction of the scattered light, and \(\hat{E}_{io}\) is the unit polarization vector of the incident probe beam’s electric field. \( N \) represents the number of electrons within this volume. \( \omega_i \) and \( \omega_s \) are the frequencies of the incident and scattered light, respectively. \( \omega \) and \( \vec{k} \) denote the frequency and wave vector of the electron density fluctuations modes that satisfy the wave–wave matching conditions:
\begin{equation}
\label{eq:match}
    \begin{aligned} 
        \vec{k}_s&=\vec{k}_i \pm \vec{k},\\
        \omega_s&=\omega_i \pm \omega,
    \end{aligned}
\end{equation}
The dynamic form factor $S(\vec{k},\omega)$ is defined by:
\begin{equation}    S(\vec{k},\omega)=\lim_{V\rightarrow\infty,T\rightarrow\infty} \frac{1}{VT}\bigg<\frac{|n_e(\vec{k},\omega)|^2}{n_{e0}}\bigg>,
\end{equation}
where \(\langle \cdot \rangle\) denotes an ensemble average, \(n_{e0}\) represents the unperturbed average electron density, \(n_e(\vec{k},\omega)\) is the electron density fluctuation spectrum, \(V\) and \(T\) are the spatial volume and temporal duration of the measurement, respectively.

Thermal CTS is one subtype of CTS and an essential diagnostic tool, corresponding to the scattering process driven by thermal fluctuations. According to the fluctuation-dissipation theorem, the corresponding spectrum in a collisionless thermal equilibrium plasma takes the form\cite{sheffield2011}:
\begin{equation}
    S(\vec{k},\omega)=\frac{2\pi}{k}\bigg|1-\frac{\chi_e}{\epsilon}\bigg|^2f_e\bigg(\frac{\omega}{k}\bigg)+\frac{2\pi Z}{k}\bigg|\frac{\chi_e}{\epsilon}\bigg|^2f_i\bigg(\frac{\omega}{k}\bigg),
\end{equation}
Where $\chi_e$ and $\chi_i$ are the susceptibilities of electron and ion respectively, $\epsilon=1+\chi_e+\chi_i$ is the plasma dielectric function, $f_e$ and $f_i$ represent the electron and ion velocity distribution function respectively\cite{fried1960}. These distributions embody key plasma parameters, including density, temperature, and flow velocity. Consequently, by fitting measured thermal CTS spectra, one can infer these plasma properties\cite{bai2001,wang2005,li2012,gong2015,zhao2019,glenzer1999,froula2006a,froula2006b,follett2016}.

However, HED plasmas under realistic ICF conditions often deviate from the collisionless thermal equilibrium assumption. Examples include high-Z collisional plasmas\cite{zheng1999}, plasmas with heat flux\cite{henchen2019}, and those exhibiting super-Gaussian electron distributions\cite{zheng1997,milder2021b}. In such regimes, conventional CTS theories become inaccurate, complicating experimental data analysis and diagnostic interpretations. This calls for new approaches to analyze how CTS spectra depend on these conditions.

Particle-in-cell (PIC) simulation is a first-principles method that self-consistently solves Maxwell’s equations and tracks particle motion under electromagnetic forces \cite{birdsall1985}, making it well-suited for studying kinetic effects in nonthermal or collisional plasmas. 
PIC simulations have been used to examine non-collective TS\cite{Zamenjani2020}. 
Simulations of electromagnetic fluctuations in pair plasmas have demonstrated good agreement with theoretical predictions \cite{ruyer2013}, validating the capability of PIC to reproduce fluctuation physics.
For CTS, where electrostatic fluctuations play a dominant role, pioneer investigations have proposed 1D and 2D PIC methods to study nonthermal plasmas \cite{Farrell2022} along certain directions.
While recent experiments seeking to obtain detailed plasma properties or to diagnose non-Maxwellian plasmas via collective Thomson scattering (CTS) often rely on collecting scattered light from multiple angles \cite{liu2019, katz2024}, high-resolution, angle- and frequency-resolved data are still required to guide experimental design.
To integrate experimental TS measurements, development of numerical simulation methods with sufficient angular and spectral resolution is essential.

Another application for CTS in ICF is the diagnosing of LPI, a key process that strongly relates with driven plasma waves. For example, in indirect-drive ICF, ion density modulations induced by the beating of multiple overlapping laser beams at the laser entrance hole of the hohlraum can significantly affect cross-beam energy transfer—a critical mechanism for implosion performance in indirect drive ICF\cite{michel2009}. Similarly, perturbations driven by the Langmuir decay instability play a pivotal role in the growth and saturation of two-plasmon-decay instability, influencing hot‑electron generation\cite{depierreux2000,follett2015,filippov2023}. The CTS measuring these externally driven plasma waves is called as super-thermal collective Thomson scattering (SCTS). Because the amplitude of driven perturbations significantly exceeds that of thermal fluctuations, SCTS signals are typically orders of magnitude stronger than thermal CTS. To support experimental studies of these processes, it is therefore essential to develop numerical methods that properly resolve angular variations in the SCTC signal under such conditions.

In this work, we developed the method to obtain CTS light spectra with high angular and spectral resolution from 2D PIC simulations, enabling the detailed analyze of  nonequilibrium anisotropic plasmas in future studies. The ion-acoustic features in the simulated spectra are clearly resolved and exhibit shapes that agree well with theoretical predictions. We further perform a systematic comparison between SCTS signals and the underlying driven density perturbations, revealing that perturbations with wave vectors mismatched from the scattering geometry still yield significant TS signals—an effect not captured by standard theory. Our analysis suggests this should result from a beating-wave interaction between the probe beam and the driven density perturbation. 

\begin{figure}
    \centering
    \includegraphics[width=0.5\linewidth]{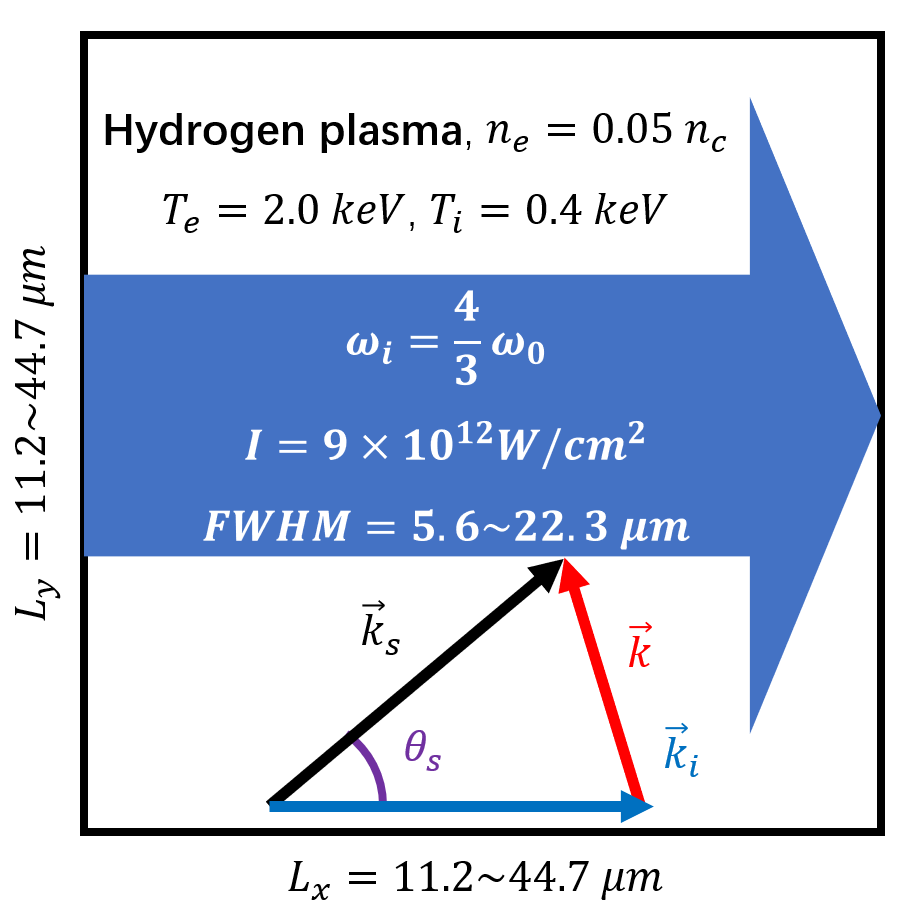}
    \caption{Scheme of PIC simulation setup used to obtain TS signal, along with the wave vector of probe beam $\vec{k}_i$, the scattered light $\vec{k}_s$ at scattering angle $\theta_s$ and the matched density perturbation $\vec{k}$.}
    \label{fig:pb-set}
\end{figure}

The remainder of this paper is organized as follows: Section \ref{sec:2_simu} describes the numerical procedure to obtain CTS signals with high resolution in angular and frequency space; Section \ref{sec:3_tcts} validates our method's ability to reproduce thermal CTS spectrum from thermal plasma consistent with current theory, including numerical results of CTS spectra with different level of ion-ion collisional effects; Section \ref{sec:4_scts} focuses on SCTS, in which the new phenomenon, drawing attention to a newly observed phenomenon in which scattered light results from perturbations with mismatched wave vectors; Section \ref{sec:5_summary} summarizes the paper and outlines directions for future work on studying ICF plasmas using TS diagnostics.

\section{SIMULATION APPROACH TO OBTAIN THOMSON SCATTERING SIGNAL}
\label{sec:2_simu}

In this section, we demonstrate the general procedure to obtain CTS signals with high frequency and angular resolutions using PIC simulation. The probe beam and plasma properties, along with temporal and spatial grid setting and sampling setting for later Fourier analysis to obtain TS signal are described. PIC simulations in this article are conducted using OSIRIS code\cite{goos2002}. An example to obtain CTS spectra in an equilibrium uniform plasma is presented.

\subsection{PIC simulation setup}
\label{sec:2.1_setup}

The simulation is performed in two dimensions (the \(xy\)-plane) on a domain ranging from \(200\,c/\omega_{0} \times 200\,c/\omega_{0}\) (\(11.2\,\mu m\times11.2\,\mu m\)) to \(800\,c/\omega_{0} \times 800\,c/\omega_{0}\) (\(44.7\,\mu m\times44.7\,\mu m\)), where \(c\) denotes the speed of light and \(\omega_{0}\) denotes the frequency of the laser beam with a wavelength of 351 nm. The domain is uniformly filled with hydrogen plasma at electron density \(n_{e} = 0.05\,n_{c}\), where \(n_{c}\) is the critical density for the 351 nm laser. The initial electron and ion temperatures are \(T_{e} = 2\,keV\) and \(T_{i} = 0.5\,keV\), respectively.

A probe beam is injected from the left boundary and propagates along the x-direction, as illustrated in Fig.\ref{fig:pb-set}. It has a frequency of \(\omega_{i} = \tfrac{4}{3}\omega_{0}\) (corresponding to a 263 nm wavelength) and is linearly polarized perpendicular to the xy-plane to maximize the CTS signal.
The transverse profile of the beam is Gaussian, with a peak intensity \(I = 9 \times 10^{12}\,\mathrm{W/cm^{2}}\) and a full width at half maximum (FWHM) varying from \(100\,c/\omega_0\) (\(5.6\,\mu m\)) to \(400\,c/\omega_0\) (\(22.3\,\mu m\)).

The spatial grid resolution is \(\Delta x = 0.25\,c/\omega_{0}\) ($0.014\,\mu m$), and the temporal step size is \(\Delta t = 0.17\,\omega_{0}^{-1}\) ($0.03\,\text{fs}$). Each grid cell contains 100 macroparticles representing electrons and another 100 for ions. Particle boundary conditions are thermal, while electromagnetic fields satisfy VPML boundary conditions\cite{vay2000} which limits EM wave reflectivity at the order of $\sim 10^{-4}$ in our setup.

\subsection{Frequency- and angular-resolved TS spectrum}
\label{sec:2.2_tfft}

To obtain the scattered-light spectrum $P_s(\theta,\omega)$ from PIC simulations, the $E_z$ field component is extracted. But the raw output $E_z(x,y,t)$ includes abundant data that can complicate analysis. To address this, OSIRIS’s time‑FFT module is applied\cite{wen2019,cao2020}, which performs in-situ Fast Fourier Transforms during the simulation and outputs only the targeted frequency range, yielding $E_z(x,y,\omega)$ and substantially reducing data volume. Provided the required resolutions in $k-$ and $\omega-$space are met, sampling in the x, y and t domain further reduces the data volume without compromising the spectral analysis.

\begin{figure}
    \centering
    \includegraphics[width=0.7\linewidth]{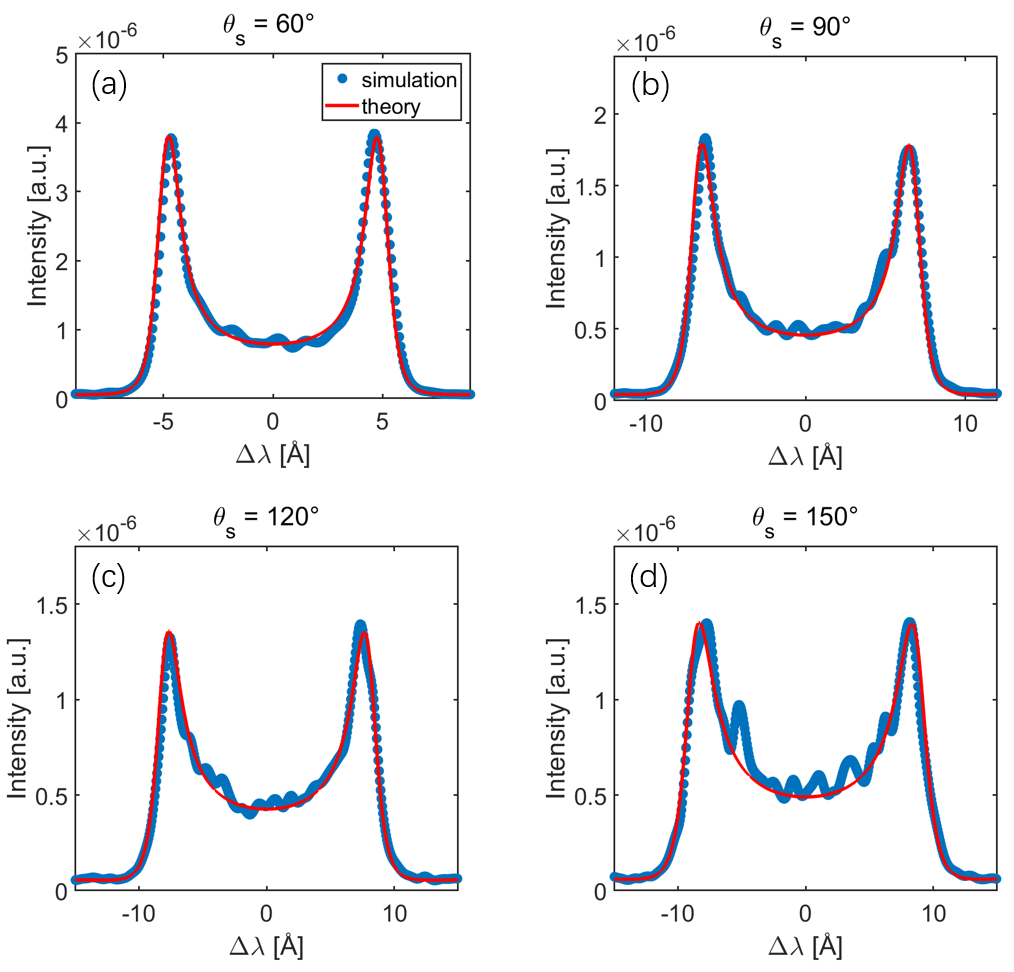}
    \caption{Thomson scattering spectra for scattering angles of $\theta_s$ of (a) $60^\circ$, (b) $90^\circ$, (c) $120^\circ$, and (d) $150^\circ$, each averaged over $\pm1^\circ$.}
    \label{fig:ThermalTS-ion}
\end{figure}

The spatial sampling interval is $4\Delta x$, and the temporal sampling interval is $2\Delta t$, corresponding to a measure range of $[-3.14,3.14]\,\omega_0/c$ for $k$ and $[-9.24,9.24]\,\omega_0$ for $\omega$. For the Thomson-scattering spectrum exhibiting ion-acoustic features generated by the $\tfrac{4}{3}\omega_{0}$ probe beam, we select the frequency window $[1.32\,\omega_{0},\,1.35\,\omega_{0}]$, which captures the relevant spectral signatures. The simulation duration is set to $175{,}000\,c/\omega_{0}$ (approximately $50\text{ ps}$), enabling a frequency resolution of $d\omega = 2.3 \times 10^{-5}\,\omega_{0}$. This value is significantly finer than the resolution typically achieved in experiments\cite{zhao2019}, where ion-acoustic features in CTS spectra are measured with a spectral resolution of about $0.06 nm$ , corresponding to $3.0\times10^{-4}\omega_0$. The wavenumber resolution is $dk = 7.8\times10^{-3}\omega_0/c$ for a simulation domain size of $800\,c/\omega_{0} \times 800\,c/\omega_{0}$ ($44.7\mu m\times44.7\mu m$).

\begin{figure}
    \centering
    \includegraphics[width=0.7\linewidth]{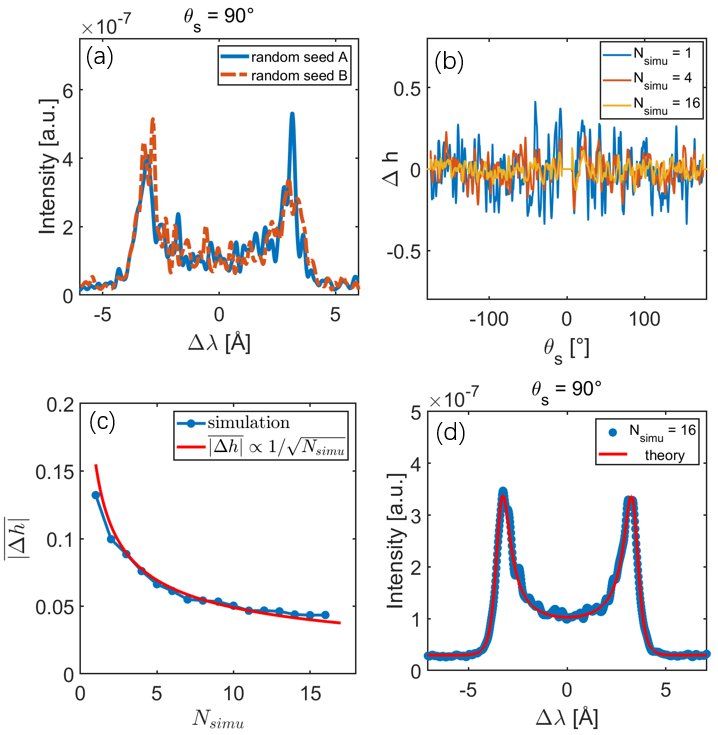}
    \caption{(a) Scattered spectra at $\theta_s=90^\circ$ from simulations with different random seed.  (b) Angular distribution of height differences from averaged results of 1, 4, 16 simulations. (c) Height difference $\overline{|\Delta h|}$ decay with the number of averaging simulations $N_{simu}$ in a manner of $1/\sqrt{N_{simu}}$. (d) Scattered spectra at $\theta_s=90^\circ$, averaged over 16 simulations.}
    \label{fig:statis}
\end{figure}

Though a spatial FFT, $E_z(x,y,\omega)$ output is transformed into $E_z(k_x,k_y,\omega)$ data. Next, a linear interpolation converts this to polar coordinates: $E_z(k,\theta,\omega)$. By applying the dispersion relation for light, $ \omega^2 = k^2c^2 + \omega_{pe}^2 $, where $\omega_{pe}$ is the plasma frequency, we obtain the angular–spectral distribution $E_z(\theta,\omega)$ for scattered light.

\section{VALIDATION: ION-ACOUSTIC FEATURES OF THERMAL THOMSON SCATTERING OBTAINED FROM SIMULATIONS}
\label{sec:3_tcts}

We apply our method to thermal CTS cases. We start from the equilibrium collision-less case, where TS spectrum can be conveniently described with existing theories.

\subsection{Ion-acoustic features of Thomson scattering}

For the plasma conditions described in Section \ref{sec:2.1_setup}, the frequency of ion-acoustic peaks \(\omega_s\) in the TS spectrum can be calculated using:
\begin{equation}
    \omega_{s}=\omega_{i}\pm\omega_{iaw}=\omega_{i}
    \pm kc_s,
\end{equation}
where $c_s\approx\sqrt{(ZT_e+3T_i)/m_i}\approx1.8\times10^{-3}c$ is the ion sound speed in this simulation, $k=k_s-k_i\approx2k_i\sin(\theta_s/2)$ is the wave number of the fluctuation matching with the scattering angle $\theta_s$. 

Following the plasma conditions in Section \ref{sec:2.1_setup} and time Fourier analysis method described in Section \ref{sec:2.2_tfft}, we obtain low-frequency part of Thomson scattering spectra at different scattering angles of \(60^\circ\pm1^\circ\), \(90^\circ\pm1^\circ\), \(120^\circ\pm1^\circ\), and \(150^\circ\pm1^\circ\) from a simulation domain with a size of \(800\,c/\omega_{0} \times 800\,c/\omega_{0}\) (\(44.7\,\mu m\times44.7\,\mu m\)), as shown in Fig.\ref{fig:ThermalTS-ion}. To reduce spectral noise, second-order smoothing with a window size of $3.0\times10^{-4}\omega_0$ is applied iteratively to these spectra until their shapes converge. 
The simulated TS spectra are in good agreement with the theoretical profiles calculated from Eq. (4), which indicates that our method can produce valid TS spectra with high angular and spectral resolution.

\subsection{Achieving higher signal-to-noise ratio}
\label{sec:snr}

Although the method described above appears straightforward, the signal-to-noise ratio presents a primary challenge in its practical application. This issue arises from the limited statistical sampling in PIC simulations. Under realistic experimental conditions, CTS spectra represent a statistical average over a three-dimensional volume of hundreds of micrometers. In contrast, PIC simulations typically operate over two-dimensional domains of only tens of micrometers. This disparity introduces much stronger fluctuations into the simulated spectra, which can significantly affect the signal-to-noise ratio of the obtained CTS spectra.

One clear example is the variation in peak heights of the ion-acoustic features, which are sensitive to the drift velocity between electrons and ions. In an isotropic plasma, the red-shifted and blue-shifted ion-acoustic peaks should be symmetric. However, in a simulation with a $200\,c/\omega_{0}\times 200\,c/\omega_{0}$ ($11.2\,\mu m\times11.2\,\mu m$) domain, the peak height asymmetry can be significant, as shown in Fig.\ref{fig:statis}(a). 

In these simulations, the initial thermal velocities of particles are randomly sampled from a Maxwellian distribution using an initial random seed. Changing this random seed does not alter macroscopic plasma quantities, such as density and temperature, but it does affect the microscopic realizations of the system. As shown in Fig.~\ref{fig:statis}(a), the pattern of height differences varies with the random seeds used to sample the initial thermal velocities, indicating that CTS spectra are affected by the initial microscopic realizations—a result that can be attributed to limited statistical sampling. We quantify this asymmetry using the normalized height difference:
\begin{equation}
    \Delta h = \frac{h_R-h_B}{h_R+h_B}
\end{equation}
where $\Delta h$ is the normalized difference, and $h_R$ and $h_B$ are the heights of the red-shifted and blue-shifted peaks, respectively. Figure~\ref{fig:statis}(b) includes the angular distribution of $\Delta h$ from a single simulation, where $\Delta h$ varies in different angles randomly. This randomness can obscure the measurement of drift velocities, thereby reducing the reliability of physical diagnostics, such as those involving electron heat transport. 

In the spectra of Fig.~\ref{fig:ThermalTS-ion}, we demonstrated that simulations with a domain size of $800\,c/\omega_{0} \times800\,c/\omega_{0}$ ($44.7\,\mu m\times44.7\,\mu m$) is adequate to yield ion-acoustic peaks of nearly equal height and a significantly improved signal-to-noise ratio. But the required simulation space may vary when different physical subjects are studied. So a more general approach to improve statistical robustness is required. This is accomplished by conducting multiple simulations under identical macroscopic conditions but with varied microscopic realizations (e.g., different random seeds for initial velocity sampling), and then averaging their resulting spectra. 

The angular distribution of $\Delta h$ for different numbers of averaged simulation results, $N_{simu}$, is shown in Fig.~\ref{fig:statis}(b), where $\Delta h$ decreases significantly as $N_{simu}$ increases. A more quantitative result is illustrated in Fig.\ref{fig:statis}(c), in which the average height difference $\overline{|\Delta h|}$ decreases following a $1/\sqrt{N_{simu}}$ scaling law. The spectrum obtained by averaging 16 simulations is shown in Fig.\ref{fig:statis}(d), where the noise level is significantly reduced compared to Fig.~\ref{fig:statis}(a).

In addition to ensemble averaging over random seeds, increasing the number of particles per cell (PPC) is a common technique for improving the signal-to-noise ratio in PIC simulations\cite{oudin2025,yin2012,yin2019}. While we have employed this method, we find it to be relatively inefficient: raising the PPC reduces both the signal and the statistical noise by similar factors, leaving the overall signal-to-noise ratio largely unchanged. Our tests show that varying the PPC by a factor of 16 results in minimal signal-to-noise ratio improvement. Further details are provided in Appendix~\ref{appn:ppc}. 
The insensitivity of signal-to-noise ratio of Thomson scattering signal to PPC could be attribute to the fact that Thomson scattering is related to the ensemble average of density perturbations, while a simulation with an increased PPC does not necessarily enhance the sampling of the fluctuation modes,
and therefore contribute little to the ensemble average.

\subsection{Capability to simulate ion-ion collision's influence on Thomson scattering spectrum}

Ion–ion collisions can modify the ion-acoustic features of the TS spectrum by enhancing collisional damping and reducing Landau damping~\cite{zheng1999}. Their influence on TS spectrum with wave vector $k$ can be evaluated with the parameter $k\lambda_{ii}$. Here $\lambda_{ii}$ is the ion mean free path due to ion-ion collisions given by $\lambda_{ii}=v_{ti}/\nu_{ii}$, where $v_{ti}$ represents ion thermal velocity and $\nu_{ii}$ denotes ion-ion collision frequency. The condition $k\lambda_{ii}\gg1$ corresponds to the collisionless cases, $k\lambda_{ii}\sim1$ to the intermediate cases, and $k\lambda_{ii}\ll1$ to the strongly collisional cases.

\begin{figure}[h]
    \centering
    \includegraphics[width=0.4\linewidth]{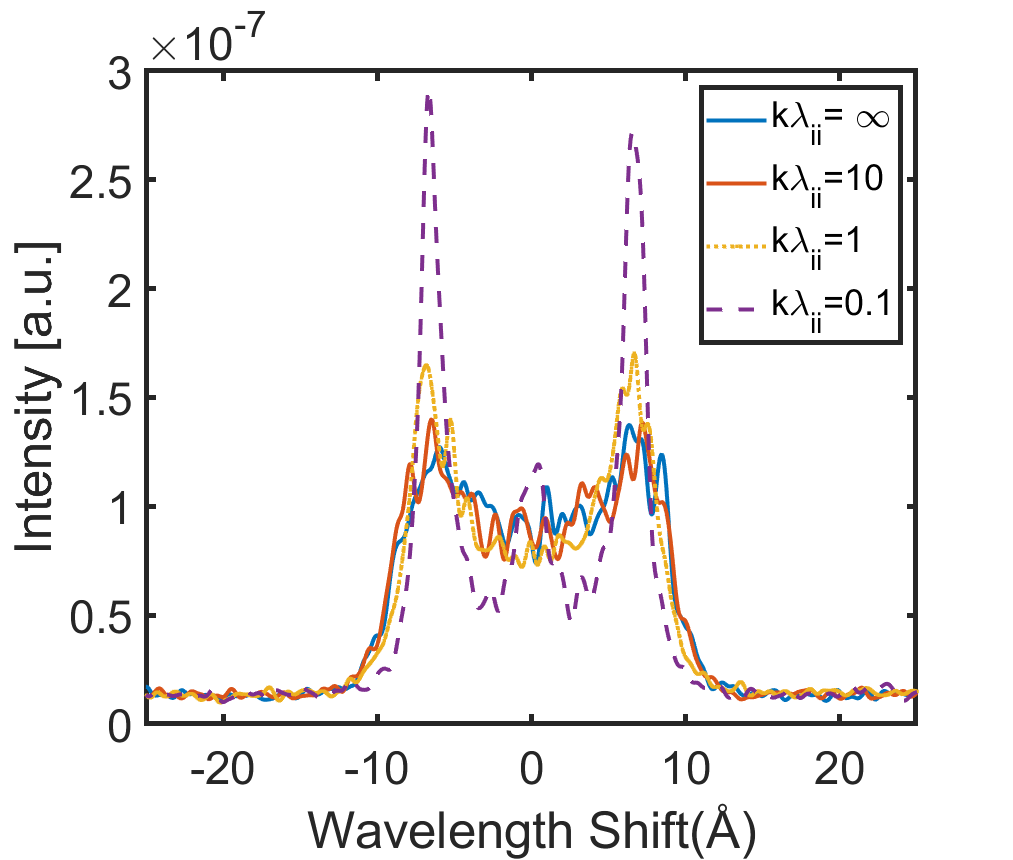}
    \caption{Simulated Thomson scattering spectra with different ion-ion collision rate.}
    \label{fig:collision}
\end{figure}

In PIC simulations, we can adjust the level of collision by setting the Coulomb logarithm manually\cite{nanbu1997,perez2012}. A series of simulations is conducted with collision parameters $k\lambda_{ii} = 0.1, 1, 10, \infty$. Here $k\lambda_{ii}=\infty$ correspond to the case where ion-ion collision is turned off in PIC simulation. Simulation domain with a size of $400\,c/\omega_0\times400\,c/\omega_0$ ($22.3\,\mu m\times22.3\,\mu m$) is filled with uniform hydrogen plasma. Electron and ion temperatures are $T_e = 2\,keV$ and $T_i=1\,keV$, respectively. Rest of parameters are the same with previous context. Simulated spectra with scattering angle of \(90^\circ\pm1^\circ\) is shown in Fig.\ref{fig:collision}.

In the collisionless case, the ion-acoustic waves undergo strong Landau damping due to $ZT_e/T_i=2$ in these simulation, rendering the resonance peaks for $k\lambda_{ii}\sim\infty$ and $k\lambda_{ii}=10$ insignificant. When $k\lambda_{ii}=1$ and $k\lambda_{ii}=0.1$, strong collisionality suppresses ion Landau damping, and the resonance peaks become pronounced. In the strongly collisional regime, an additional resonance peak at zero frequency shift, corresponding to an entropy wave, appears. These tendencies in the simulated spectra are consistent with the theoretical predictions~\cite{zheng1999}, confirming that the simulation properly reproduces the effect of collisions.

In addition, in the above work we intentionally retain only ion-ion collisions to enable a direct and uncluttered comparison with existing theoretical models that specifically account for this process. This approach allows us to cleanly demonstrate the capability of our PIC-based method to reproduce CTS spectra under varying collision strengths. However, the ion-electron and electron-electron collisions should also influence the CTS spectrum. As noted in prior work \cite{Epperlein1992,Berger2005}, ion-electron collisions modify the damping rate of ion acoustic waves (IAWs). Similarly, electron-electron collisions can affect the electron contribution to IAW Landau damping. Under conditions where ion-ion collisions are significant, these additional processes may also become non-negligible.
While a comprehensive treatment of all collision types is beyond the scope of this initial methodological study, a detailed investigation of multi-species collisional impacts on CTS is planned for future work.

\section{SUPER-THERMAL THOMSON SCATTERING VIA PIC SIMULATION}
\label{sec:4_scts}

Unlike thermal CTS, SCTS arises from light scattering off driven plasma waves whose amplitudes far exceed thermal fluctuations, making it less susceptible to signal-to-noise limitations in PIC simulations. In such simulations, the amplitude, wave vector, and frequency of the driven waves can be precisely controlled, enabling reproduction of the SCTS process and facilitating a targeted study of the matching condition in Eq.~(2).

\begin{figure}[h]
\centering
\includegraphics[width=0.7\linewidth]{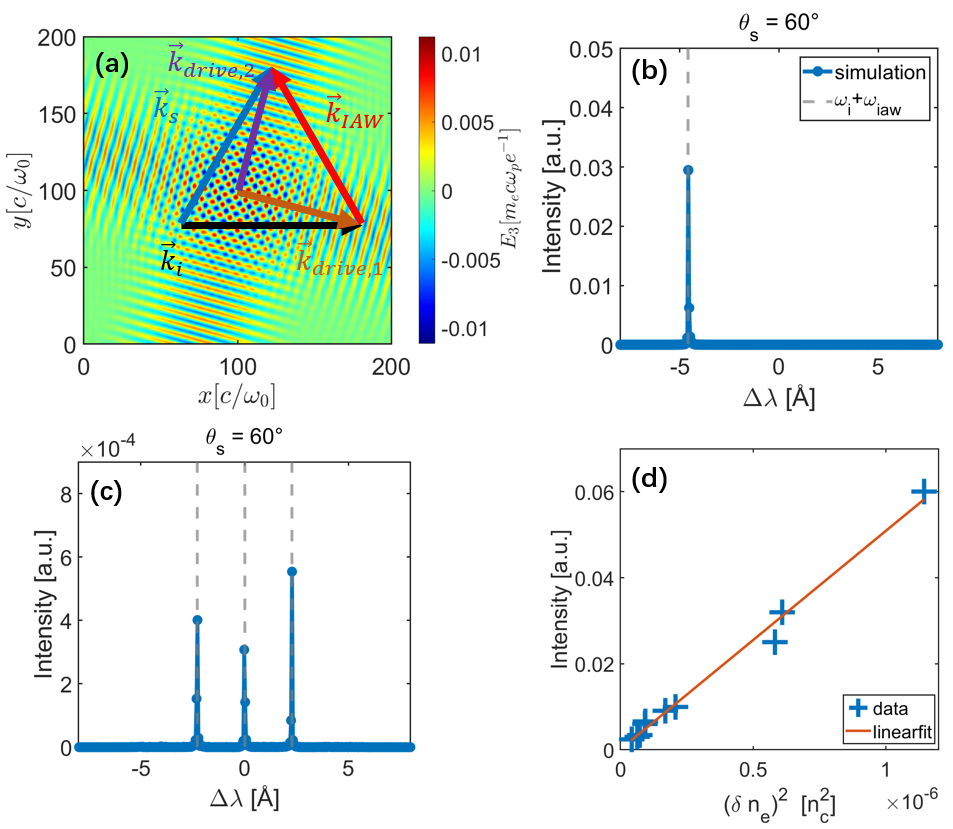}
\caption{(a) Two laser beams beating to drive an ion density perturbation. (b) TS spectrum at $60^\circ$ with SCTS peak corresponding to the driven perturbation frequency. (c) TS spectrum of three perturbations with different frequencies but the same wave vector. (d) Relation between the SCTS power and the amplitude of the corresponding perturbation.}
\label{fig:STS-A}
\end{figure}

In this section, we show that when the matching condition is well satisfied, the frequency distribution of the simulated SCTS signal reflects the frequency of the driven perturbation, while its angular distribution reflects the perturbation’s wave vector. We also identify a new phenomenon in which scattered light originates from density perturbations with a mismatched wave vector.

\subsection{Super-thermal Thomson scattering from matched density perturbations with different frequency}

To investigate SCTS induced by matched density perturbations, we generate these perturbations through laser beating. When two lasers $(\vec{k}_{drive,1},\omega_{drive,1})$ and $(\vec{k}_{drive,2},\omega_{drive,2})$ overlap spatially, their interference envelope has a wave vector and frequency of $(\vec{k}_{drive,1}-\vec{k}_{drive,2},\omega_{drive,1}-\omega_{drive,2})$. The ponderomotive force associated with this interference pattern drives a density fluctuation. By adjusting the wave vectors and frequencies of the beams, density perturbations with arbitrary wave vector and frequency can be generated.

The simulations are performed in a domain of $200\,c/\omega_0 \times 200\,c/\omega_0$ ($11.2\, \mu m\times11.2\,\mu m$), filled with uniform hydrogen plasma. The plasma and probe parameters are the same as those described in Section \ref{sec:2.1_setup}. As shown in Fig.\ref{fig:STS-A}(a), two laser beams, each with an intensity of $3.5\times10^{14}\,\mathrm{W/cm^2}$ and a wave vector of $0.93\,\omega_0/c$, overlap at an angle of $90^\circ$, driving an IAW with a wave vector $k_{\mathrm{iaw}}$ matched to $60^\circ$ Thomson scattering. The frequency difference between the two driving beams corresponds to the IAW eigenfrequency, $\omega_{\mathrm{iaw}}=k_{\mathrm{iaw}}c_s$. The corresponding SCTS signal at $60^\circ$ is shown in Fig.~\ref{fig:STS-A}(b), where a narrow-band peak appears at the matched frequency $\omega_i+\omega_{\mathrm{iaw}}$.

We further excite three ion density perturbations with the same wave vector but different frequencies in a single simulation. The configuration is the same as in Fig.\ref{fig:STS-A}(a), except that one drive beam $(\vec{k}_{\mathrm{drive},1},\omega_{\mathrm{drive}}+\omega_{\mathrm{iaw}})$ is replaced by three overlapping beams with frequencies $\omega_{\mathrm{drive}}-\tfrac{1}{2}\omega_{\mathrm{iaw}}$, $\omega_{\mathrm{drive}}$, and $\omega_{\mathrm{drive}}+\tfrac{1}{2}\omega_{\mathrm{iaw}}$, respectively. The intensity of each drive beam is set to $1\times10^{14}\,\mathrm{W/cm^2}$. Three SCTS signals are simultaneously observed in the $60^\circ$ scattering spectrum, with frequency shifts corresponding to the driven perturbations, as shown in Fig.\ref{fig:STS-A}(c).

\begin{figure}
    \centering
    \includegraphics[width=0.7\linewidth]{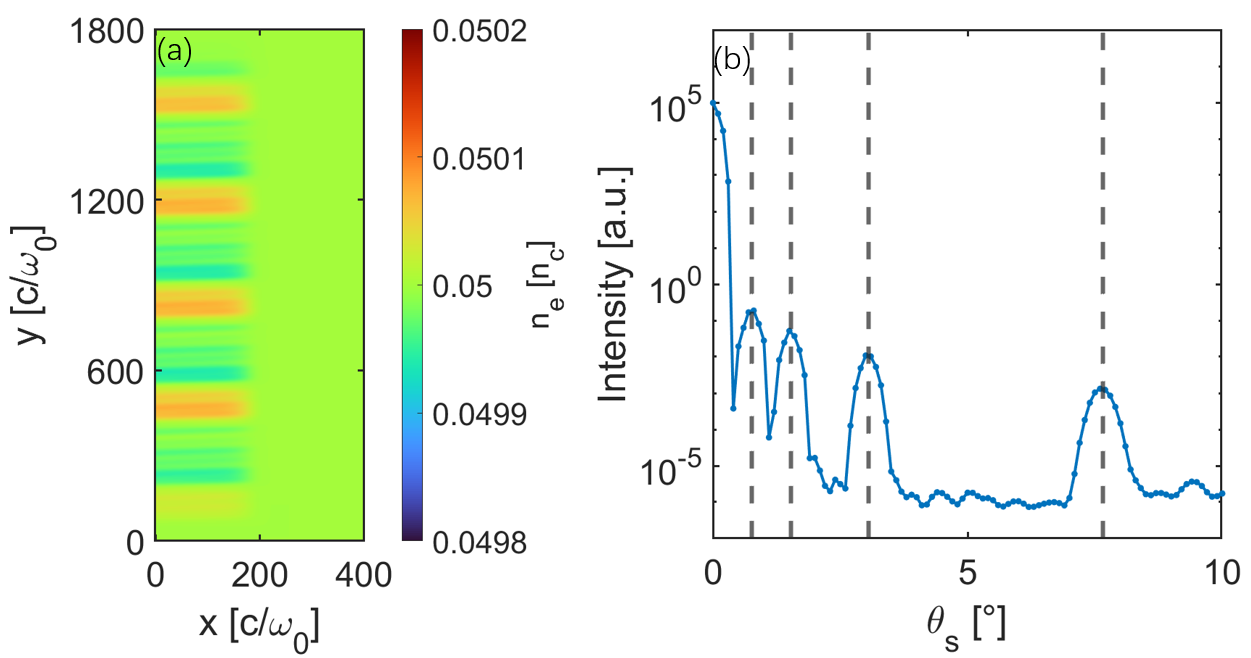}
    \caption{(a) Initial electron density distribution with four density perturbations at the left half of simulation domain. (b) Angular distribution of zero-frequency-shifted signal. The blue solid line is the distribution in simulation result, the gray dashed lines are the theoretical position of SCTS peaks.}
    \label{fig:STS-B}
\end{figure}

A series of simulations is performed using the same configuration described above, but with varying drive-beam intensities and frequency differences. These variations generate density perturbations of different amplitudes, leading to corresponding changes in the SCTS signals. By comparing the amplitudes of the SCTS signals with the perturbation amplitudes, we find that the scattered power is proportional to the square of the perturbation amplitude, as shown in Fig.~\ref{fig:STS-A}(d). This result indicates that the frequency distribution of the TS signal directly reflects the frequency distribution of the density perturbations when the matching condition is satisfied.

\subsection{Super-thermal Thomson scattering from matched density perturbations with different wave vector}

When density perturbations are driven by laser beating, the region of laser overlap constitutes only a small fraction of the total simulation area. This leads to inefficient use of computational resources. Furthermore, to generate multiple perturbation modes, the laser beating method requires a correspondingly greater number of laser beams. This approach often excites more modes than are necessary, introducing unwanted complexity.
For typical cases associated with static density perturbations, an alternative method is to initialize the electrons with a perturbed density distribution \(n_e(x,y)|_{t=0}\)
without any ions injected into the simulation domain. Owing to the nature of PIC simulation~\cite{birdsall1985}, it automatically enforces the Poisson equation at any time $t$:
\begin{equation}
    \label{eq:poisson}
    \nabla\cdot\vec{E}-4\pi\rho=C,
\end{equation}
where $C$ is a constant determined by initial conditions. 
When the simulation is initialized with $E|_{t=0}=0$ and \(\rho|_{t=0}=-en_e(x,y)|_{t=0}\), we obtain $C=4\pi n_e(x,y)|_{t=0}$. This form implies that, at any time \(t\), the electric potential corresponds to that generated by the instantaneous electron density superimposed on a stationary background ion density equal to \(n_e(x,y)|_{t=0}\).

To investigate SCTS arising from matched density perturbations with different wave vectors, we initialize four electron density fluctuations within a single simulation, as shown in Fig.~\ref{fig:STS-B}(a). The simulation is conducted in a domain of $400\,c/\omega_0 \times1800\,c/\omega_0$ ($22.4\,\mu\mathrm{m}\times100.6\,\mu\mathrm{m}$), filled with a background electron density of $n_0=0.05n_c$. The boundaries perpendicular to the $y$-axis are set to be periodic, while the remaining boundaries were thermal for particles and open for electromagnetic fields.

The density perturbations are initialized in the left half of the simulation domain (see Fig.~\ref{fig:STS-B}(a)). These perturbations have wavelengths of $20\,\mu\mathrm{m}$, $10\,\mu\mathrm{m}$, $5\,\mu\mathrm{m}$ and $2\,\mu\mathrm{m}$, each oriented such that its wave vector matched that of Thomson-scattered light at a different scattering angle. Their amplitudes are set to $(1,0.5,0.25,0.1)\times10^{-3}n_0$, respectively.

A plane wave with frequency $\omega_{i}=\tfrac{4}{3}\omega_{0}$ is launched from the left boundary along the $x$-axis as a probe beam. The scattered signal is collected in the right half of the simulation domain. This configuration, which separates the interaction region from the detection region, is designed to distinguish far-field electromagnetic signals from near-field ones and will also be discussed in the next section. The angular distribution of the TS signal is shown in Fig.~\ref{fig:STS-B}(b), where SCTS features are observed at different angles that agree with those predicted by the matching condition. This result demonstrates that the TS spectrum effectively represents the wave vector distribution of matched electron density fluctuations.

Moreover, the angular intensity distributions of the scattered light shown in Fig.~\ref{fig:STS-B}(b) have finite widths for each scattering angles, with the widths increasing with the scattered angles. 
This reveals the presence of scattered signals at angles slightly deviating from the exact matching condition. 
Such scattering under wave-vector-mismatched conditions can be partially attributed to the finite wave-vector broadening of both the probe beam and the driven density perturbation, which relaxes the strict wave-vector-matching requirement by allowing a finite range of wave-vector components to overlap with the matching condition. 
However, the angular distribution becomes increasingly broader at larger scattering angles, suggesting that wave-vector broadening alone cannot fully account for the observed signal, as analyzed in Appendix~\ref{appn:Dtheta}. 
This behavior therefore points to the presence of additional scattering mechanisms operating under wave-vector-mismatched conditions, which are discussed in Section~\ref{sec:4.3_mismatch}.

\subsection{Thomson scattering from perturbations with a wave vector deviated from matching condition}
\label{sec:4.3_mismatch}

The SCTS under matching conditions has been presented in the previous sections. 
When measuring driven plasma waves using SCTS, the scattering geometry should be carefully designed to satisfy the matching condition. 
However, plasma conditions and the resulting driven plasma waves are often difficult to predict, and the matching conditions may not always be fulfilled in experiments. 

To investigate a representative scenario relevant to ICF plasmas, where Thomson scattering is employed to diagnose laser-beat–driven ion waves, we perform a PIC simulation to examine this process under a mismatched condition.
The simulation is performed in two dimensions (the \(xy\)-plane) on a domain measuring \(400\,c/\omega_{0} \times 400\,c/\omega_{0}\) (\(22.4\,\mu m\times22.4\,\mu m\)). The domain is uniformly filled with a CH plasma at an electron density of \(n_{e} = 0.056\,n_{c}\). Initial electron and ion temperatures are set to \(T_{e} = 2\,\text{keV}\) and \(T_{i} = 1\,\text{keV}\), respectively. The spatial resolution is \(\Delta x = 0.25\,c/\omega_{0}\) (\(0.014\,\mu\text{m}\)), and the temporal step is \(\Delta t = 0.17\,\omega_{0}^{-1}\) (\(0.03\,\text{fs}\)). Each grid cell is populated with 100 macroparticles for electrons and 50 for each ion species. Particle boundary conditions are thermal, while electromagnetic fields satisfy VPML boundary conditions\cite{vay2000}.

\begin{figure}
    \centering
    \includegraphics[width=0.9\linewidth]{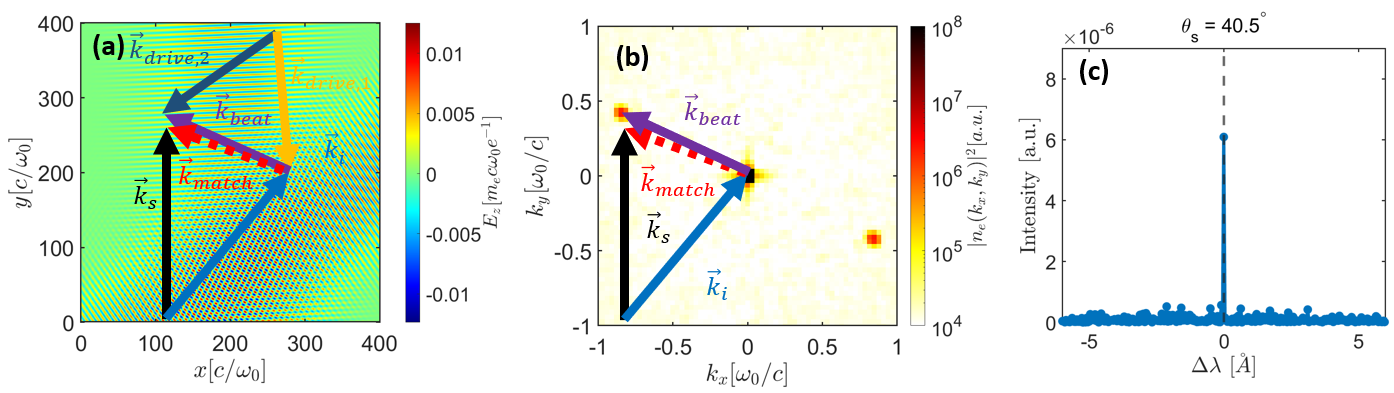}
    \caption{(a) Distribution of electric field $E_z$ in the simulation domain, along with the wave vectors of drive beams [$\vec{k}_{drive,1}$ (orange) and $\vec{k}_{drive,2}$ (dark blue)], the driven density perturbation [$\vec{k}_{beat}$ (purple)], the probe beam [$\vec{k}_i$ (blue)], the scattered light [$\vec{k}_s$ (black)] and the matched density perturbation [$\vec{k}_{match}$ (dashed red)]. (b) The spatial spectrum of the electron density perturbation in the phase space $(k_x,k_y)$. (c) The SCTS spectrum along $\vec{k}_s$ with the peak position marked by the vertical dashed line.}
    \label{fig:stsDevi}
\end{figure}

The simulation employs two drive beams, each with frequency \(\omega_0\), which beat against each other as illustrated in Fig.~\ref{fig:stsDevi}(a). For the plasma with \(n_e=0.056\,n_c\), the electron plasma frequency is \(\omega_{pe}=0.23\omega_0\). The electromagnetic dispersion relation yields a wave number \(k = 0.97\,\omega_0/c\) for frequency \(\omega_0\). As shown in the geometry of Fig.~\ref{fig:stsDevi}(a), drive beam 1 has a wave vector \(\vec{k}_{\text{drive},1} \approx (0.04,-0.97)\omega_0/c\), oriented \(2.5^\circ\) counter-clockwise from the negative \(y\)-axis. Drive beam 2 has \(\vec{k}_{\text{drive},2} \approx (-0.81,-0.54)\omega_0/c\), oriented \(33.5^\circ\) counter-clockwise from the negative \(x\)-axis. The wave vector of the resulting laser-beating density perturbation is \(\vec{k}_{\text{beat}} \approx (-0.85,0.43)\omega_0/c\).

A probe beam with frequency \(\omega_i = 4\omega_0/3\) is launched at an angle of \(40.5^\circ\) clockwise from the positive \(y\)-axis. The polarization of the probe beam is set perpendicular to the simulation plane to maximize the Thomson scattering signal. The matched wave vector \(\vec{k}_{\text{match}}\) is derived from the frequency-matching condition \(\omega_s = \omega_i + \omega_{\text{beat}} = 4\omega_0/3\), where the beat frequency \(\omega_{\text{beat}} = 0\). For the probe frequency of \(4\omega_0/3\), the magnitude of the wave vector is \(|\vec{k}_i| \approx 1.31\,\omega_0/c\). Specifically, the probe's wave vector is \(\vec{k}_i \approx (0.85,0.99)\omega_0/c\). Light scattered along the \(y\)-axis has a wave vector \(\vec{k}_s \approx (0,1.31)\omega_0/c\). The matched wave vector is therefore \(\vec{k}_{\text{match}} = \vec{k}_s - \vec{k}_i \approx (-0.85,0.32)~\omega_0/c\). The resulting wave-vector mismatch is \(\Delta \vec{k} = \vec{k}_{\text{beat}} - \vec{k}_{\text{match}} \approx (0,0.11)\omega_0/c\), which corresponds to \(0.08\,|\vec{k}_s|\).

The corresponding density perturbation distribution displayed in Fig.~\ref{fig:stsDevi}(b) shows no signal at the matched position. However, in the scattering spectrum shown in Fig.~\ref{fig:stsDevi}(c), a SCTS signal with an amplitude significantly higher than thermal TS signal is observed. 
The wave vector broadening at finite beam waist is analyzed in Appendix~\ref{appn:kbroad}, and this broadening does not fully account for the observed signal.
This result indicates that the observable scattered light reflects not only the amplitude of the matched mode but also contributions from surrounding unmatched modes. This introduces additional challenges in interpreting SCTS signals from driven ion waves.

\begin{figure}
    \centering
    \includegraphics[width=1.0\linewidth]{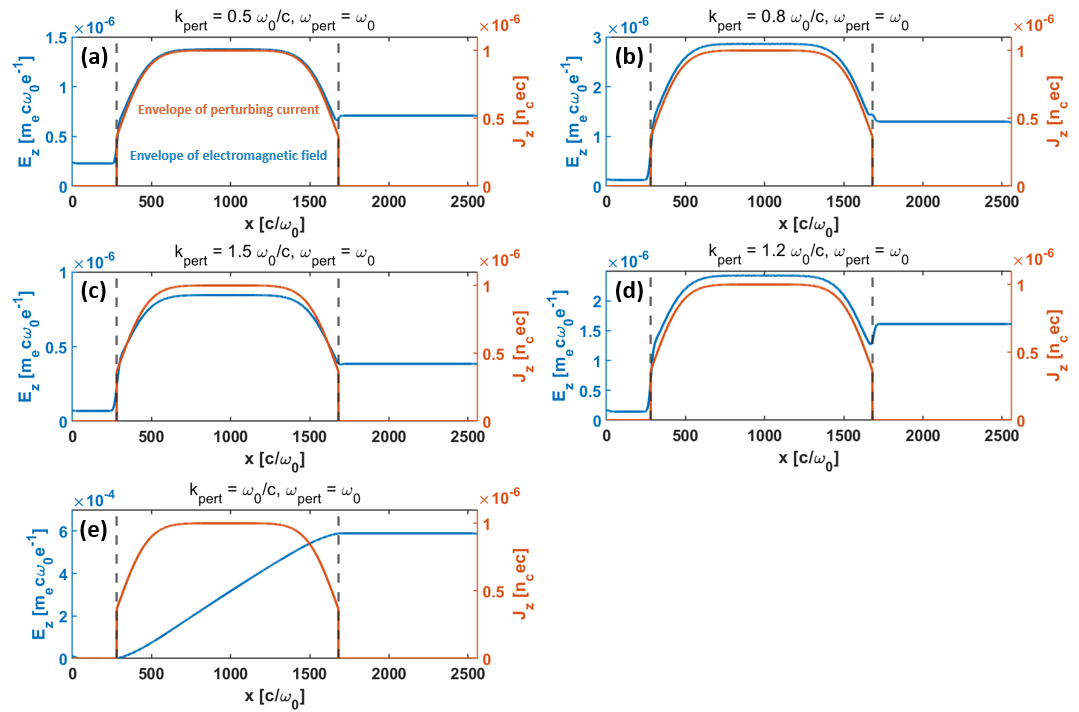}
    \caption{
    Spatial envelopes of the perturbed current density $\vec J_{\rm pert}$ (red), along with the envelopes of excited electromagnetic field (blue) for (a) $\omega_{\rm pert} = \omega_0$, $k_{\rm pert} = 0.5 \omega_0/c$, (b) $\omega_{\rm pert} = \omega_0$, $k_{\rm pert} = 0.8 \omega_0/c$, (c) $\omega_{\rm pert} = \omega_0$, $k_{\rm pert} = 1.5 \omega_0/c$, (d) $\omega_{\rm pert} = \omega_0$, $k_{\rm pert} = 1.2 \omega_0/c$ and (e) $\omega_{\rm pert} = \omega_0$, $k_{\rm pert} = \omega_0/c$.
    }
    \label{fig:STS-C}
\end{figure}

The occurrence of SCTS from density perturbation modes that do not satisfy the conventional wave-vector matching conditions [Eq.~(\ref{eq:match})] appears to contradict the standard theory of Thomson scattering, which requires strict frequency and wave-vector matching between the probe beam, the scattered light and the density perturbation. 
To resolve this puzzle, we must return to the fundamental principle of electromagnetic (EM) wave emission from oscillating electrons.
Irrespective of whether the matching conditions are met, the total emitted EM field should always be a coherent superposition of the waves radiated by all oscillating currents induced by the electron motion. 
Consequently, an analysis of the radiation produced by this current distribution is essential to uncover the underlying physics of SCTS from unmatched modes.

We consider the interaction of the probe beam electric field $\vec{E}_i\sin(\vec{k}_i\cdot \vec{x}-\omega_it)$ with electron density perturbation $\delta n_e\sin(\vec{k}_{0}\cdot \vec{x}-\omega_{0}t)$. 
A current density perturbation can be directly generated via wave beating: 
\begin{equation}
\begin{aligned} 
    \vec{J}_{pert}&=\frac{\vec{E}_ie\delta n_e}{m_e}\sin(\vec{k}_i\cdot \vec{x}-\omega_it)\sin(\vec{k}_{0}\cdot \vec{x}-\omega_{0}t)\\
    &=\frac{\vec{E}_ie\delta n_e}{2m_e}(\cos[(\vec{k}_i-\vec{k}_{0})\cdot \vec{x}-(\omega_i-\omega_{0})t]-\cos[(\vec{k}_i+\vec{k}_{0})\cdot \vec{x}-(\omega_i+\omega_{0})t]).
\end{aligned}
\end{equation}
This $\vec{J}_{pert}$, with wave vector $\vec{k}_{pert}=\vec{k}_i\pm\vec{k}_{0}$ and frequency $\omega_{pert}=\omega_i\pm\omega_{0}$, could excite EM waves\cite{matsukiyo2016}. Starting from the Maxwell equation, we obtain the relation between currents and electric fields:
\begin{equation}
\label{eq:JtoE1}
    \nabla\times(\nabla\times \vec{E})+\frac{1}{c^2}\frac{\partial^2 \vec{E}}{\partial t^2}=-\frac{4\pi}{c^2}\frac{\partial \vec{J}}{\partial t},
\end{equation}
in which $\vec{J}$ includes $\vec{J}_{pert}$, the perturbation current driven by the interaction between the probe beam and the density perturbations, and $\vec{J}_{cond}$, the conduction current arising from the interaction of background electrons with the electric field. For high frequency electromagnetic wave in a non-magnetized plasma, $\vec{J}_{cond}$ satisfies $\frac{\partial \vec{J}_{cond}}{\partial t}=\frac{e^2n_e}{m_e}\vec{E}$. Substitute $\vec{J}=\vec{J}_{pert}+\vec{J}_{cond}$ into Eq. (\ref{eq:JtoE1}), we get:
\begin{equation}
\nabla\times(\nabla\times \vec{E})+\frac{1}{c^2}\frac{\partial^2 \vec{E}}{\partial t^2}=-\frac{\omega_{pe}^2}{c^2}\vec{E} -\frac{4\pi}{c^2}\frac{\partial \vec{J}_{pert}}{\partial t}.
\end{equation}
Then we have:
\begin{equation}
\label{eq:JtoE2}
(\nabla^2-\frac{1}{c^2}\frac{\partial^2}{\partial t^2}-\frac{\omega_{pe}^2}{c^2})\vec{E} = \frac{4\pi}{c^2}\frac{\partial \vec{J}_{pert}}{\partial t}.
\end{equation}

Equation~(\ref{eq:JtoE2}) describes the growth of an EM wave driven by a perturbative current density, \(\vec{J}_{\mathrm{pert}}\), in a plasma with frequency $\omega_{pe}$. 
If the wave vector \(\vec{k}_{\mathrm{pert}}\) and frequency \(\omega_{\mathrm{pert}}\) satisfy the EM dispersion relation of the background plasma, the matching condition of Eq.~(\ref{eq:match}) is fulfilled. 
In this resonant case, the emission is strong because the driven wave corresponds to an eigenmode of the system.
Conversely, when the matching condition is not satisfied, the driven wave is not an eigenmode and cannot propagate over long distances in the plasma.
Nevertheless, such non-eigenmodes can still give rise to scattered radiation of eigenmodes with finite amplitude.
Physically, as these driven non-eigenmodes exit the interaction region between the probe beam and the density perturbations, their oscillating EM fields act as boundary sources that can excite propagating EM eigenmodes at the same frequency in the outer regions.

We demonstrate this mechanism—the excitation of eigenmodes by localized non-eigenmodes—using PIC simulations.
It is important to note that, while Eq.~\eqref{eq:JtoE2} describes electromagnetic emission in a plasma, the underlying mechanism is more general and does not inherently depend on the presence of a plasma.
To isolate and illustrate the essential physics, we therefore perform the demonstration in vacuum, where no physical particles are present in the simulation domain.
In this simplified setup, the perturbative current \(\vec{J}_{\mathrm{pert}}\) is imposed directly as an external source.
Thus, we insert the following external electron current into a fully vacuum one-dimensional PIC simulation:
\begin{equation}
\vec{J}_{\rm{pert}}=\vec{J}_0\sin(k_{\rm{pert}}x-\omega_{\rm{pert}}t)\exp\left[-(\frac{x-x_0}{L_{\rm{pert}}})^6\right],
\end{equation}
where $L_{\rm pert} = 700\,c/\omega_0$ is the length of the perturbation region, and $x_0 = 980\,c/\omega_0$ represents its center in the simulation domain. We set the current to zero beyond the width $L_{pert}$.

This 1D model is a simplified setup to explicitly reveal the emission mechanism. The scenarios map directly with the 2D simulations: the imposed current \(J_{\mathrm{pert}}\) in 1D represents the source from the probe–perturbation interaction in 2D. The mismatch in Fig.~\ref{fig:STS-C}(a-d) corresponds to choosing \(\omega_{\mathrm{pert}}/k_{\mathrm{pert}} \neq c\) (i.e., violating the EM dispersion relation), analogous to the \(\Delta \vec{k}\) mismatch in Fig.~\ref{fig:stsDevi}(b). Please note that the amplitude of $\vec{J}_{\rm pert}$ is not crucial to the primary aim of the simulations presented in Fig.~(\ref{fig:STS-C}), which is to demonstrate whether emission occurs under matched or mismatched conditions.

Figure (\ref{fig:STS-C}) shows the spatial envelope of $\vec{J}_{\rm pert}$ and the resulting electromagnetic field for five cases with $\omega_{\rm pert} = \omega_0$ and $k_{\rm pert}$ = 0.5, 0.8, 1.5, 1.2, 1.0 $\omega_0/c$. In the case corresponding to matched wave vector ($k_{\rm pert}=1.0$), a propagating electromagnetic wave is excited within the perturbed region and radiates outward [Fig.~\ref{fig:STS-C}(e)]. 
By contrast, other cases with mismatched wave vectors feature a near-field electromagnetic perturbation localized to the driven region; at the boundary, 
scattered light at frequency $\omega_{pert}$ is emitted into the surrounding vacuum. These results indicate that scattered light can still be generated even under unmatched wave-vector.
Moreover, We observe that the intensity of the scattered light is highest when \(\omega_{\mathrm{pert}} / k_{\mathrm{pert}}\) approaches the light speed \(c\), corresponding to the matched condition. The emission amplitude decreases systematically as the ratio deviates from \(c\), consistent with the expected behavior for off-resonant driving.

Additionally, we point out that the primary purpose of the vacuum simulation is to isolate and illustrate the core physical mechanism of eigenmode excitation by a localized non-eigenmode source, not to model the full plasma dispersion quantitatively. While this simplification will inevitably lead to quantitative deviations in the wave properties (e.g., a slight shift in the effective wave number for a given frequency), it does not alter the fundamental process by which a non-propagating field excites a propagating eigenmode at a boundary.

Therefore, since driven perturbations that deviate from the matching condition can still produce SCTS signals, the actual ion modes underlying the diagnosed SCTS signals should not be restricted to the exact matching condition. This needs to be carefully investigated and taken care of in analyzing our experimental results on Shenguang facility. 

\section{Summary}
\label{sec:5_summary}

In summary, we have developed a first-principles numerical approach for obtaining scattered light signals of ion acoustic features with high angular and frequency resolution under typical ICF conditions, using PIC simulations. 
The method reproduces existing theoretical predictions for thermal collective Thomson scattering, and extends to the super-thermal regime where driven plasma modes are present.
In cases where the driven modes are well-matched in wave vectors with the probe and collection geometries, the results agree well with theoretical expectations. 
Importantly, we find that significant TS signals can persist even under imperfect matching conditions, in contrast to the conventional knowledge that the TS spectrum strictly follows the plasma density spectrum. 
This discrepancy is explained by a beating-wave mechanism arising from the interaction between the probe beam and the driven plasma density modulations. These findings highlight that SCTS signals can carry contributions from both matched and unmatched modes, and thus provide a practical framework for interpreting scattering spectra in the presence of driven ion modes, which are a common yet complex feature in ICF plasmas.

In this work, our analysis has focused on the ion-acoustic features of the Thomson scattering spectrum, which are typically the dominant components under ICF conditions. However, the electron feature is also an important diagnostic target in experiments. Because its intensity is several orders of magnitude lower than that of the ion-acoustic feature, accurately resolving it in PIC simulations remains challenging due to statistical noise. 
In future work, we plan to employ our simulation method to obtain electron features. We also aim to apply the present framework to more specific ICF-relevant problems, such as diagnosing CBET-related density perturbations in laser entrance hole region of hohlraums using SCTS.

\ack{
This work was supported by the National Key R\&D Program of China, No. 2023YFA1608400 and National Natural Science Foundation of China (NSFC) (Grant No. 12275269, U2430207, 12275251). The numerical simulations in this paper were conducted on Hefei advanced computing center. We thank the UCLA-IST OSIRIS Consortium for the use of OSIRIS. 
}

\appendix

\section{Influence of Particle-per-Cell Resolution on Signal-to-Noise Ratio}
\label{appn:ppc}

\begin{figure}[h]
    \centering
    \includegraphics[width=1.0\linewidth]{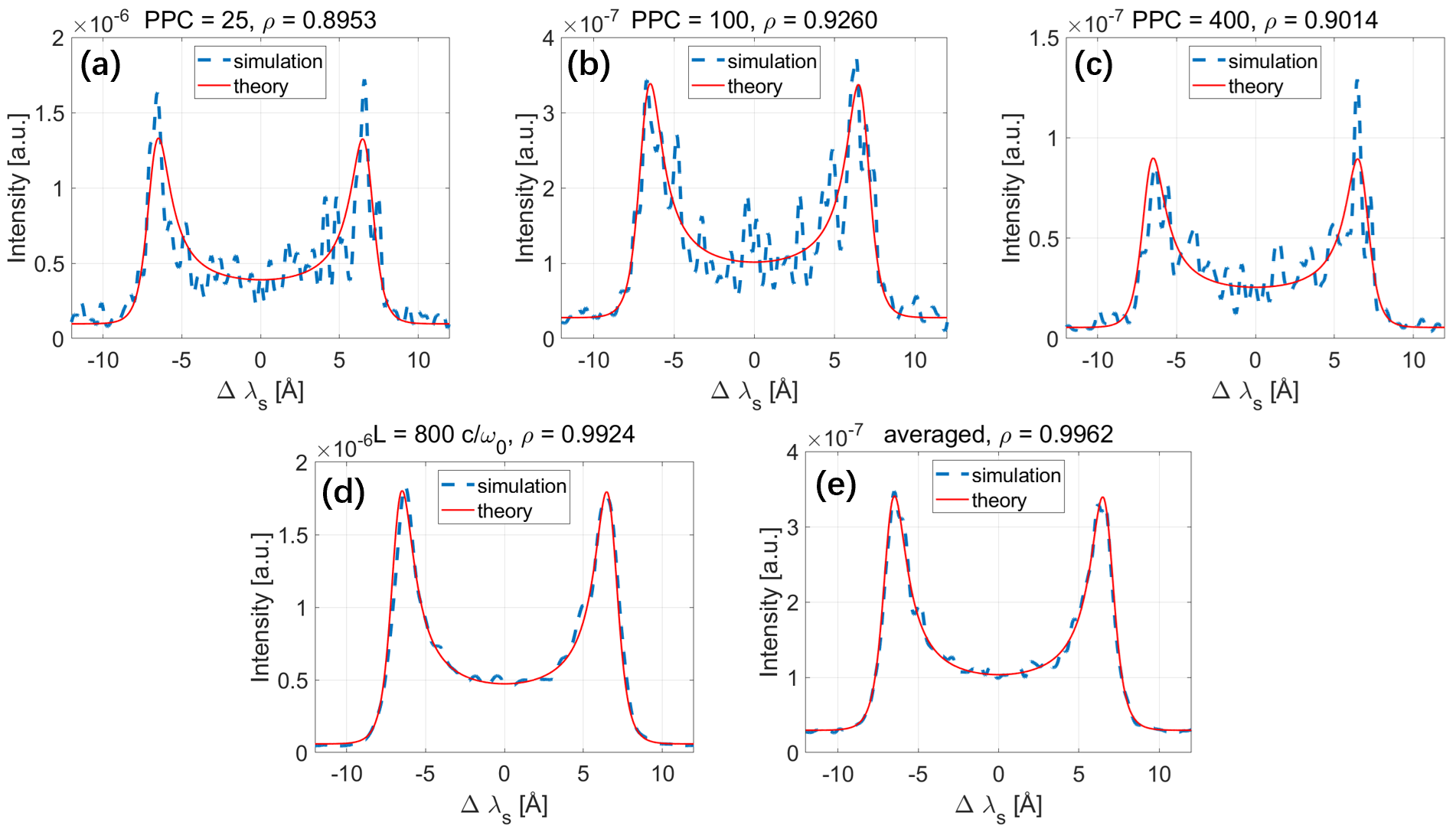}
    \caption{Simulated (blue dashed line) and theoretical (red solid line) scattering spectra at $\theta_s = 90^\circ$ along with the correlation coefficient $\rho$ between them. (a–c): single simulations with domain size $200c/\omega_0 \times 200c/\omega_0$ and PPC = 25, 100, and 400, respectively. (d): single simulation with enlarged domain $800c/\omega_0 \times 800c/\omega_0$ and PPC = 100. (e): ensemble-averaged spectra over 16 independent simulations with domain $200c/\omega_0 \times 200c/\omega_0$ and PPC = 100.}
    \label{fig:ppc}
\end{figure}

To evaluate the effect of particle-per-cell (PPC) resolution on the thermal collective Thomson scattering spectra, we performed PIC simulations with 25, 100, and 400 macroparticles per cell for both electrons and ions. All other parameters were identical to those used in Fig.~\ref{fig:statis}(a). The resulting spectra at a scattering angle $\theta_s = 90^\circ$ are shown in Fig.~\ref{fig:ppc}(a-c).

The absolute amplitude of the simulated CTS signal depends on PPC because the fluctuation level in PIC simulations is influenced by finite macroparticle statistics. Increasing PPC reduces the discrete particle noise, leading to a decrease in both the background fluctuation level and the scattered spectral amplitude. Therefore, the absolute signal intensity from an individual simulation is not a strictly converged quantity.

In contrast, the normalized spectral shape—which is the primary diagnostic observable—remains consistent across the tested PPC values. The characteristic double-peak structure is preserved, and the agreement with the theoretical spectrum, quantified by the correlation coefficient $\rho$ (Fig.~\ref{fig:ppc}(a–c)), changes only marginally with increasing PPC. Although higher PPC reduces statistical noise, a single simulation remains a stochastic realization with finite sampling of fluctuation modes, and thus the ensemble-averaged spectrum is not fully represented.

Improved convergence is achieved either by enlarging the simulation domain or by ensemble averaging over independent runs. As shown in Fig.~\ref{fig:ppc}(d) and (e), both a domain enlarged by a factor of 16 and the average of 16 independent simulations yield correlation coefficients exceeding 0.99. 

We therefore employ ensemble averaging to obtain high-fidelity CTS spectra within the adopted simulation framework, while a quantitative comparison of absolute spectral intensities with experimental measurements is beyond the scope of this work.

\section{Influence of Wave Vector Broadening on Angular Broadening of Scattered Light}
\label{appn:Dtheta}

\begin{figure}[h]
    \centering
    \includegraphics[width=0.75\linewidth]{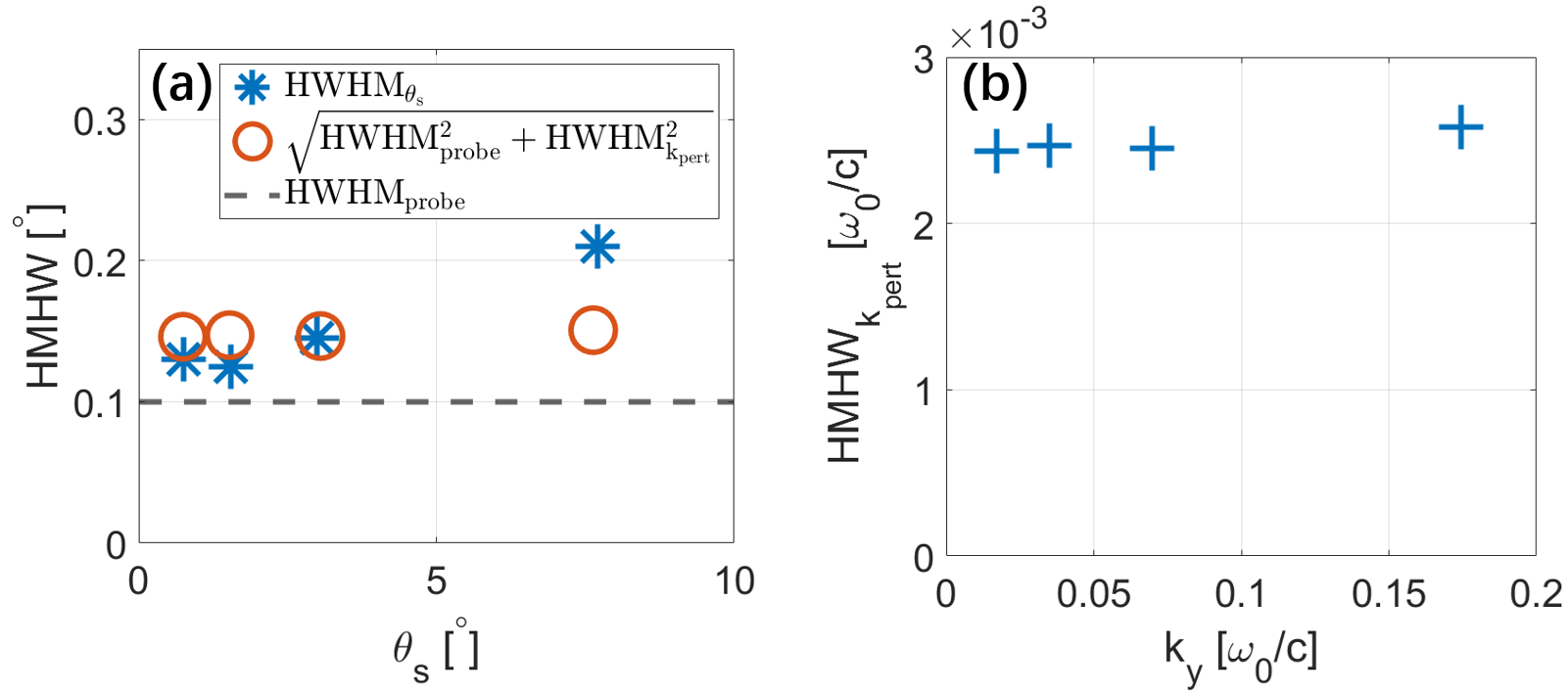}
    \caption{(a) Half-width at half-maximum (HWHM) of the angular distribution as a function of scattering angle (blue asterisks). The gray dashed line denotes the HWHM of the probe beam. Red circles denote the quadrature sum of the probe-beam HWHM and the HWHM associated with wave-vector broadening of the density perturbations. (b) HWHM of wave vector spectrum of electron density perturbation corresponding to different scattered light.}
    \label{fig:Dtheta}
\end{figure}

The angular distribution of the scattered light shown in Fig.~\ref{fig:STS-B}(b) exhibits finite peak widths for both the probe beam and the scattered signal. To quantify the role of wave-vector broadening in determining the angular spread of the scattered light, we analyze the dependence of the scattered angular width on the spectral broadening of both the probe beam and the driven density perturbation. As shown in Fig.~\ref{fig:Dtheta}(a), the half-width at half-maximum (HWHM) of the scattered light exceeds that of the probe beam, indicating that the wave-vector broadening of the density perturbation must also be taken into account.

In Fig.~\ref{fig:Dtheta}(b), the wave-vector broadening $\Delta k_{\rm pert}$ of the density perturbation at different $k_y$ is evaluated using the HWHM of the spectral peak. After considering the widths of both the probe beam and density modulations via $\sqrt{\Delta \theta_{probe}^2+\Delta \theta_{pert}^2}$ with $\Delta \theta_{pert} = \frac{\Delta k_{pert}}{k_{i}}$, the total contributions are plotted in Fig.~\ref{fig:Dtheta}(a) by red circles. We find that these results cannot explain the growing width of the scattered lights for greater $\theta_s$. These results demonstrate that although wave-vector broadening contributes to the angular spread of the scattered signal, it is insufficient to fully account for the scattering observed under wave-vector-mismatched conditions.

\section{Influence of Wave Vector Broadening on Mismatched Scattering}
\label{appn:kbroad}

In realistic Thomson scattering configurations, the interaction occurs within a finite spatial region of characteristic length $L$, leading to a finite wave-vector broadening of order $1/L$. 
Under wave-vector-mismatched conditions, such finite spectral widths of the probe beam and the driven density perturbation could in principle allow the scattering to happen if the imposed wave vector deviation lies within the effective spectral width.

\begin{figure}[h]
    \centering
    \includegraphics[width=1\linewidth]{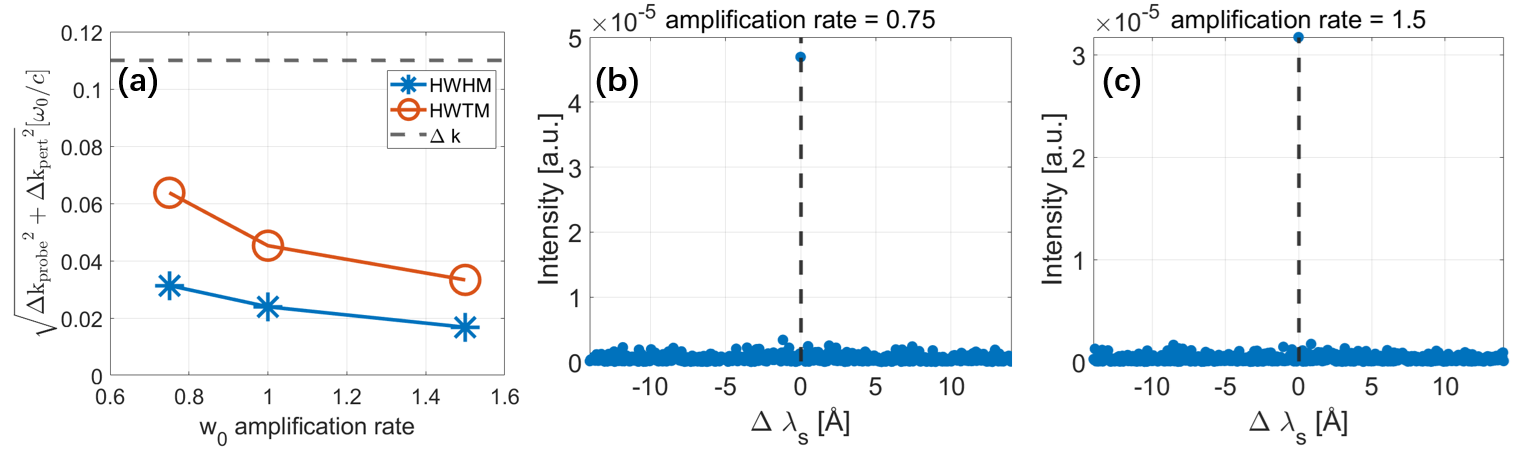}
    \caption{(a) Half width at half maximum (HWHM) of the quadrature sum of the probe-beam and density-perturbation broadenings along with the wave vector deviation from matched condition $\Delta k$. (b–c) Scattered spectra at $40.5^\circ$ for scaling factors of 0.75 and 1.5, respectively.}
    \label{fig:deltak}
\end{figure}

To test whether the observed mismatched scattering can be attributed solely to finite spectral broadening, we varied the waist of both the probe and drive beams in the simulation corresponding to Fig.~\ref{fig:stsDevi}, using scaling factors of 0.75 and 1.5 relative to the reference case. 
The corresponding wave-vector broadenings and scattered spectra are shown in Fig.~\ref{fig:deltak}.

To evaluate the combined effect of these broadenings on the scattering process, we estimate an effective spectral width given by the quadrature sum $\sqrt{{\Delta k_{probe}}^2+{\Delta k_{pert}}^2}$. As shown in Fig.~\ref{fig:deltak}(a), even in the smallest-waist case, the combined HWHM (blue asterisks) and HWTM (red circles) remain substantially smaller than $|\Delta k|$ (gray dashed line). Nevertheless, clear scattering persists in all cases, as shown in Fig.~\ref{fig:deltak}(b) and (c), indicating that this scattering under wave vector mismatched condition cannot be fully explained by the finite wave vector broadening.


\begin{thebibliography}{10}

\bibitem{langdon1980}
A.~Bruce Langdon.
\newblock Nonlinear {Inverse} {Bremsstrahlung} and {Heated}-{Electron} {Distributions}.
\newblock {\em Physical Review Letters}, 44(9):575--579, March 1980.

\bibitem{turnbull2020}
David Turnbull, Arnaud Colaïtis, Aaron~M. Hansen, Avram~L. Milder, John~P. Palastro, Joseph Katz, Christophe Dorrer, Brian~E. Kruschwitz, David~J. Strozzi, and Dustin~H. Froula.
\newblock Impact of the {Langdon} effect on crossed-beam energy transfer.
\newblock {\em Nature Physics}, 16(2):181--185, February 2020.

\bibitem{turnbull2024}
D.~Turnbull, J.~Katz, M.~Sherlock, A.~L. Milder, M.~S. Cho, L.~Divol, N.~R. Shaffer, D.~J. Strozzi, P.~Michel, and D.~H. Froula.
\newblock Reconciling calculations and measurements of inverse bremsstrahlung absorption.
\newblock {\em Physics of Plasmas}, 31(6):063304, June 2024.

\bibitem{luciani1983}
J.~F. Luciani, P.~Mora, and J.~Virmont.
\newblock Nonlocal {Heat} {Transport} {Due} to {Steep} {Temperature} {Gradients}.
\newblock {\em Physical Review Letters}, 51(18):1664--1667, October 1983.

\bibitem{henchen2019}
R.~J. Henchen, M.~Sherlock, W.~Rozmus, J.~Katz, P.~E. Masson-Laborde, D.~Cao, J.~P. Palastro, and D.~H. Froula.
\newblock Measuring heat flux from collective {Thomson} scattering with non-{Maxwellian} distribution functions.
\newblock {\em Physics of Plasmas}, 26(3):032104, March 2019.

\bibitem{michel2009}
P.~Michel, L.~Divol, E.~A. Williams, S.~Weber, C.~A. Thomas, D.~A. Callahan, S.~W. Haan, J.~D. Salmonson, S.~Dixit, D.~E. Hinkel, M.~J. Edwards, B.~J. MacGowan, J.~D. Lindl, S.~H. Glenzer, and L.~J. Suter.
\newblock Tuning the {Implosion} {Symmetry} of {ICF} {Targets} via {Controlled} {Crossed}-{Beam} {Energy} {Transfer}.
\newblock {\em Physical Review Letters}, 102(2):025004, January 2009.

\bibitem{kirkwood2013}
R~K Kirkwood, J~D Moody, J~Kline, E~Dewald, S~Glenzer, L~Divol, P~Michel, D~Hinkel, R~Berger, E~Williams, J~Milovich, L~Yin, H~Rose, B~MacGowan, O~Landen, M~Rosen, and J~Lindl.
\newblock A review of laser–plasma interaction physics of indirect-drive fusion.
\newblock {\em Plasma Physics and Controlled Fusion}, 55(10):103001, October 2013.
\newblock Number: 10.

\bibitem{craxton2015}
R.~S. Craxton, K.~S. Anderson, T.~R. Boehly, V.~N. Goncharov, D.~R. Harding, J.~P. Knauer, R.~L. McCrory, P.~W. McKenty, D.~D. Meyerhofer, J.~F. Myatt, A.~J. Schmitt, J.~D. Sethian, R.~W. Short, S.~Skupsky, W.~Theobald, W.~L. Kruer, K.~Tanaka, R.~Betti, T.~J.~B. Collins, J.~A. Delettrez, S.~X. Hu, J.~A. Marozas, A.~V. Maximov, D.~T. Michel, P.~B. Radha, S.~P. Regan, T.~C. Sangster, W.~Seka, A.~A. Solodov, J.~M. Soures, C.~Stoeckl, and J.~D. Zuegel.
\newblock Direct-drive inertial confinement fusion: {A} review.
\newblock {\em Physics of Plasmas}, 22(11):110501, November 2015.
\newblock Number: 11.

\bibitem{li2023}
X.~F. Li, S.~M. Weng, P.~Gibbon, H.~H. Ma, S.~H. Yew, Z.~Liu, Y.~Zhao, M.~Chen, Z.~M. Sheng, and J.~Zhang.
\newblock Transition from backward to sideward stimulated {Raman} scattering with broadband lasers in plasmas.
\newblock {\em Matter and Radiation at Extremes}, 8(6):065601, November 2023.

\bibitem{wang2023}
Qiang Wang, Zhichao Li, Zhanjun Liu, Tao Gong, Wenshuai Zhang, Tao Xu, Bin Li, Ping Li, Xin Li, Chunyang Zheng, Lihua Cao, Xincheng Liu, Kaiqiang Pan, Hang Zhao, Yonggang Liu, Bo~Deng, Lifei Hou, Yingjie Li, Xiangming Liu, Yulong Li, Xiaoshi Peng, Zanyang Guan, Qiangqiang Wang, Xingsen Che, Sanwei Li, Qiang Yin, Wei Zhang, Liqiong Xia, Peng Wang, Xiaohua Jiang, Liang Guo, Qi~Li, Minqing He, Liang Hao, Hongbo Cai, Wudi Zheng, Shiyang Zou, Dong Yang, Feng Wang, Jiamin Yang, Baohan Zhang, Yongkun Ding, and Xiantu He.
\newblock The effects of incident light wavelength difference on the collective stimulated {Brillouin} scattering in plasmas.
\newblock {\em Matter and Radiation at Extremes}, 8(5):055602, September 2023.

\bibitem{hao2023}
Liang Hao, Jie Qiu, and Wen~Yi Huo.
\newblock Generation of high intensity speckles in overlapping laser beams.
\newblock {\em Matter and Radiation at Extremes}, 8(2):025903, March 2023.

\bibitem{liu2024}
Q.~K. Liu, L.~Deng, Q.~Wang, X.~Zhang, F.~Q. Meng, Y.~P. Wang, Y.~Q. Gao, H.~B. Cai, and S.~P. Zhu.
\newblock Electron kinetic effects in back-stimulated {Raman} scattering bursts driven by broadband laser pulses.
\newblock {\em Matter and Radiation at Extremes}, 9(4):047402, July 2024.

\bibitem{lian2025}
C.-W. Lian, Y.~Ji, R.~Yan, J.~Li, L.-F. Wang, Y.-K. Ding, and J.~Zheng.
\newblock Two-plasmon-decay instability stimulated by dual laser beams in inertial confinement fusion.
\newblock {\em Matter and Radiation at Extremes}, 10(1):017403, January 2025.

\bibitem{sheffield2011}
John Sheffield, editor.
\newblock {\em Plasma scattering of electromagnetic radiation: experiment, theory and computation}.
\newblock Elsevier, Amsterdam ; Boston, 1st ed edition, 2011.

\bibitem{fried1960}
B~D Fried, M~Gell-Mann, J~D Jackson, and H~W Wyld.
\newblock Longitudinal plasma oscillations in an electric field.
\newblock {\em Journal of Nuclear Energy. Part C, Plasma Physics, Accelerators, Thermonuclear Research}, 1(4):190--198, January 1960.

\bibitem{bai2001}
Bo~Bai, Jian Zheng, Wandong Liu, C.~X. Yu, Xiaohua Jiang, Xiaodong Yuan, Wenhong Li, and Z.~J. Zheng.
\newblock Thomson scattering measurement of gold plasmas produced with 0.351 $\mu$m laser light.
\newblock {\em Physics of Plasmas}, 8(9):4144--4148, September 2001.

\bibitem{wang2005}
Zhebin Wang, Jian Zheng, Bin Zhao, C.~X. Yu, Xiaohua Jiang, Wenhong Li, Shenye Liu, Yongkun Ding, and Zhijian Zheng.
\newblock Thomson scattering from laser-produced gold plasmas in radiation conversion layer.
\newblock {\em Physics of Plasmas}, 12(8), August 2005.

\bibitem{li2012}
Zhichao Li, Jian Zheng, Xiaohua Jiang, Zhebin Wang, Dong Yang, Huan Zhang, Sanwei Li, Qiang Yin, Fanghua Zhu, Ping Shao, Xiaoshi Peng, Feng Wang, Liang Guo, Peng Yuan, Zheng Yuan, Li~Chen, Shenye Liu, Shaoen Jiang, and Yongkun Ding.
\newblock Interaction of 0.53 $\mu$m laser pulse with millimeter-scale plasmas generated by gasbag target.
\newblock {\em Physics of Plasmas}, 19(6), June 2012.

\bibitem{gong2015}
Tao Gong, Zhichao Li, Xiaohua Jiang, Yongkun Ding, Dong Yang, Zhebin Wang, Fang Wang, Ping Li, Guangyue Hu, Bin Zhao, Shenye Liu, Shaoen Jiang, and Jian Zheng.
\newblock Development of {Thomson} scattering system on {Shenguang}-{III} prototype laser facility.
\newblock {\em Review of Scientific Instruments}, 86(2):023501, February 2015.

\bibitem{zhao2019}
Hang Zhao, Zhichao Li, Dong Yang, Xin Li, Yaohua Chen, Xiaohua Jiang, Yonggang Liu, Tao Gong, Liang Guo, Sanwei Li, Qi~Li, Feng Wang, Shenye Liu, Jiamin Yang, Shaoen Jiang, Wanguo Zheng, Baohan Zhang, and Yongkun Ding.
\newblock Progress in optical {Thomson} scattering diagnostics for {ICF} gas-filled hohlraums.
\newblock {\em Matter and Radiation at Extremes}, 4(5):055201, September 2019.

\bibitem{glenzer1999}
S.~H. Glenzer, W.~E. Alley, K.~G. Estabrook, J.~S. De~Groot, M.~G. Haines, J.~H. Hammer, J.-P. Jadaud, B.~J. MacGowan, J.~D. Moody, W.~Rozmus, L.~J. Suter, T.~L. Weiland, and E.~A. Williams.
\newblock Thomson scattering from laser plasmas.
\newblock {\em Physics of Plasmas}, 6(5):2117--2128, May 1999.

\bibitem{froula2006a}
D.~H. Froula, J.~S. Ross, L.~Divol, N.~Meezan, A.~J. MacKinnon, R.~Wallace, and S.~H. Glenzer.
\newblock Thomson-scattering measurements of high electron temperature hohlraum plasmas for laser-plasma interaction studies.
\newblock {\em Physics of Plasmas}, 13(5):052704, May 2006.
\newblock Number: 5.

\bibitem{froula2006b}
D.~H. Froula, J.~S. Ross, L.~Divol, and S.~H. Glenzer.
\newblock Thomson-scattering techniques to diagnose local electron and ion temperatures, density, and plasma wave amplitudes in laser produced plasmas (invited).
\newblock {\em Review of Scientific Instruments}, 77(10):10E522, October 2006.

\bibitem{follett2016}
R.~K. Follett, J.~A. Delettrez, D.~H. Edgell, R.~J. Henchen, J.~Katz, J.~F. Myatt, and D.~H. Froula.
\newblock Plasma characterization using ultraviolet {Thomson} scattering from ion-acoustic and electron plasma waves (invited).
\newblock {\em Review of Scientific Instruments}, 87(11):11E401, November 2016.

\bibitem{zheng1999}
Jian Zheng, C.~X. Yu, and Z.~J. Zheng.
\newblock The dynamic form factor for ion-collisional plasmas.
\newblock {\em Physics of Plasmas}, 6(2):435--443, February 1999.

\bibitem{zheng1997}
Jian Zheng, C~X Yu, and Z~J Zheng.
\newblock Effects of non-{Maxwellian} (super-{Gaussian}) electron velocity distribution on the spectrum of {Thomson} scattering.
\newblock {\em Phys. Plasmas}, 4(2736), 1997.

\bibitem{milder2021b}
A~L Milder, J~Katz, R~Boni, J~P Palastro, M~Sherlock, W~Rozmus, and D~H Froula.
\newblock Measurements of {Non}-{Maxwellian} {Electron} {Distribution} {Functions} and {Their} {Effect} on {Laser} {Heating}.
\newblock {\em PHYSICAL REVIEW LETTERS}, 2021.

\bibitem{birdsall1985}
Charles~K. Birdsall and A.~Bruce Langdon.
\newblock {\em Plasma physics via computer simulation}.
\newblock McGraw-Hill, New York, 1985.

\bibitem{ruyer2013}
C.~Ruyer, L.~Gremillet, D.~Bénisti, and G.~Bonnaud.
\newblock Electromagnetic fluctuations and normal modes of a drifting relativistic plasma.
\newblock {\em Physics of Plasmas}, 20(11):112104, November 2013.

\bibitem{Zamenjani2020}
F.~{Moradi Zamenjani}, M.~{Ali Asgarian}, M.~Mostajaboddavati, and C.~Rasouli.
\newblock Particle-in-cell simulation feasibility test for analysis of non-collective thomson scattering as a diagnostic method in iter.
\newblock {\em Nuclear Engineering and Technology}, 52(3):568--574, 2020.

\bibitem{Farrell2022}
Audrey Farrell, Chaojie Zhang, Yipeng Wu, Zan Nie, Noa Nambu, Mitchell Sinclair, Kenneth Marsh, and Chandrashekhar Joshi.
\newblock Thomson scattering diagnostics of nonthermal plasma from particle-in-cell simulations.
\newblock In {\em 2022 IEEE Advanced Accelerator Concepts Workshop (AAC)}, pages 1--6, 2022.

\bibitem{liu2019}
Yaoyuan Liu, Yongkun Ding, and Jian Zheng.
\newblock Improvement in {Thomson} scattering diagnostic precision via fitting the multiple-wavenumber spectra simultaneously.
\newblock {\em Review of Scientific Instruments}, 90(8):083501, August 2019.

\bibitem{katz2024}
J.~Katz, R.~Boni, A.~L. Milder, D.~Nelson, K.~Daub, and D.~H. Froula.
\newblock Measurement of {Thomson}-scattering spectra with continuous angular resolution (invited).
\newblock {\em Review of Scientific Instruments}, 95(9):093513, September 2024.

\bibitem{depierreux2000}
S.~Depierreux, C.~Labaune, J.~Fuchs, and H.~A. Baldis.
\newblock Application of {Thomson} scattering to identify ion acoustic waves stimulated by the {Langmuir} decay instability.
\newblock {\em Review of Scientific Instruments}, 71(9):3391--3401, September 2000.

\bibitem{follett2015}
R.~K. Follett, D.~H. Edgell, R.~J. Henchen, S.~X. Hu, J.~Katz, D.~T. Michel, J.~F. Myatt, J.~Shaw, and D.~H. Froula.
\newblock Direct observation of the two-plasmon-decay common plasma wave using ultraviolet {Thomson} scattering.
\newblock {\em Physical Review E}, 91(3):031104, March 2015.

\bibitem{filippov2023}
E.~D. Filippov, M.~Khan, A.~Tentori, P.~Gajdos, A.~S. Martynenko, R.~Dudzak, P.~Koester, G.~Zeraouli, D.~Mancelli, F.~Baffigi, L.~A. Gizzi, S.~A. Pikuz, Ph.D. Nicolaï, N.~C. Woolsey, R.~Fedosejevs, M.~Krus, L.~Juha, D.~Batani, O.~Renner, and G.~Cristoforetti.
\newblock Characterization of hot electrons generated by laser–plasma interaction at shock ignition intensities.
\newblock {\em Matter and Radiation at Extremes}, 8(6):065602, November 2023.

\bibitem{goos2002}
R.~A. Fonseca, L.~O. Silva, F.~S. Tsung, V.~K. Decyk, W.~Lu, C.~Ren, W.~B. Mori, S.~Deng, S.~Lee, T.~Katsouleas, and J.~C. Adam.
\newblock {OSIRIS}: {A} {Three}-{Dimensional}, {Fully} {Relativistic} {Particle} in {Cell} {Code} for {Modeling} {Plasma} {Based} {Accelerators}.
\newblock In Gerhard Goos, Juris Hartmanis, Jan Van~Leeuwen, Peter M.~A. Sloot, Alfons~G. Hoekstra, C.~J.~Kenneth Tan, and Jack~J. Dongarra, editors, {\em Computational {Science} — {ICCS} 2002}, volume 2331, pages 342--351. Springer Berlin Heidelberg, Berlin, Heidelberg, 2002.
\newblock Series Title: Lecture Notes in Computer Science.

\bibitem{vay2000}
Jean-Luc Vay.
\newblock A {New} {Absorbing} {Layer} {Boundary} {Condition} for the {Wave} {Equation}.
\newblock {\em Journal of Computational Physics}, 165(2):511--521, December 2000.

\bibitem{wen2019}
H.~Wen, A.~V. Maximov, R.~Yan, J.~Li, C.~Ren, and F.~S. Tsung.
\newblock Three-dimensional particle-in-cell modeling of parametric instabilities near the quarter-critical density in plasmas.
\newblock {\em Physical Review E}, 100(4):041201, October 2019.

\bibitem{cao2020}
S.~H. Cao, R.~Yan, H.~Wen, J.~Li, and C.~Ren.
\newblock Cogeneration of hot electrons from multiple laser-plasma instabilities.
\newblock {\em Physical Review E}, 101(5):053205, May 2020.

\bibitem{oudin2025}
A.~Oudin, Y.~Lalaire, G.~Bouchard, A.~Debayle, A.~Fusaro, P.~Loiseau, C.~Ruyer, and D.~Benisti. 
\newblock heory and simulations of cross-beam energy transfer between speckled laser beams.
\newblock {\em Physics of Plasmas}, 32(4):042706, April 2025.

\bibitem{yin2012}
L.~Yin, B.~J. Albright, H.~A. Rose, K.~J. Bowers, B.~Bergen, R.~K. Kirkwood, D.~E. Hinkel, A.~B. Langdon, P.~Michel, D.~S. Montgomery, and J.~L. Kline.
\newblock Trapping induced nonlinear behavior of backward stimulated Raman scattering in multi-speckled laser beams.
\newblock {\em Physics of Plasmas}, 19(5):056304, May 2012.

\bibitem{yin2019}
L.~Yin, B.~J. Albright, E.~L. Vold, W.~D. Nystrom, R.~F. Bird, and K.~J. Bowers.
\newblock Plasma kinetic effects on interfacial mix and burn rates in multispatial dimensions.
\newblock {\em Physics of Plasmas}, 26(6):062302, June 2019.

\bibitem{nanbu1997}
K.~Nanbu.
\newblock Theory of cumulative small-angle collisions in plasmas.
\newblock {\em Physical Review E}, 55(4):4642--4652, April 1997.

\bibitem{perez2012}
F.~Pérez, L.~Gremillet, A.~Decoster, M.~Drouin, and E.~Lefebvre.
\newblock Improved modeling of relativistic collisions and collisional ionization in particle-in-cell codes.
\newblock {\em Physics of Plasmas}, 19(8):083104, August 2012.

\bibitem{Epperlein1992}
E.~M. Epperlein, R.~W. Short, and A.~Simon.
\newblock {Damping of ion-acoustic waves in the presence of electron-ion collisions}.
\newblock {\em Physical Review Letters}, 69(12):1765--1768, 1992.

\bibitem{Berger2005}
R.~L. Berger and E.~J. Valeo.
\newblock {The frequency and damping of ion acoustic waves in collisional and collisionless two-species plasma}.
\newblock {\em Physics of Plasmas}, 12(3):032104, 2005.

\bibitem{matsukiyo2016}
S~Matsukiyo, Y~Kuramitsu, and K~Tomita.
\newblock Collective scattering of an incident monochromatic circularly polarized wave in an unmagnetized non-equilibrium plasma.
\newblock {\em Journal of Physics: Conference Series}, 688:012062, March 2016.

\end{thebibliography}
\end{document}